\documentstyle[aps,epsf,bbold]{revtex}

\newcommand{\beq}{\begin{equation}}
\newcommand{\eq}{\end{equation}}
\newcommand{\beqa}{\begin{eqnarray}}
\newcommand{\eqa}{\end{eqnarray}}
\begin{document}
\preprint{MKPH-T-00-06}
\title{%
Two-body effects in coherent $\eta$-meson photoproduction on the deuteron 
in the region of the $S_{11}(1535)$ resonance\thanks{Supported
by the Deutsche Forschungsgemeinschaft (SFB 443)}}
\author{Frank Ritz and Hartmuth Arenh{\"o}vel}

\address{Institut f{\"u}r Kernphysik, Johannes Gutenberg-Universit{\"a}t, 
D-55099 Mainz, Germany}
\date{\today}

\maketitle

\begin{abstract}
Coherent $\eta$-meson photoproduction on the deuteron 
has been studied, where the emphasis is on the relative importance of 
two-body contributions from hadronic rescattering and electromagnetic 
meson exchange currents besides the impulse approximation.
For the elementary 
photoproduction amplitude a coupled resonance model developed by Bennhold 
and Tanabe has been used which fits reasonably well the experimental data.
The rescattering effects are treated within a coupled
channel approach considering the intermediate excitation of the 
$P_{11}(1440)$, $D_{13}(1520)$, and $S_{11}(1535)$ nucleon resonances.
The hadronic interaction between nucleon and resonances is modeled by one 
boson exchange potentials, which we have considered both in the static 
approximation as well as fully retarded. The sum of all considered two-body 
effects results in an enhancement of the total cross section between 10 in 
the maximum and 25 percent closer to threshold around 680 MeV 
if the hadronic interaction is treated 
retarded. This enhancement shows up in the differential cross sections 
mainly at backward angles. It increases steadily from only a few percent at 
$0^\circ$ to more than a factor two at $180^\circ$ for a photon energy of 
680 MeV. 
Two-body effects become also significant in certain 
polarization observables. Finally, no discrepancy has been found for the 
ratio of the isoscalar amplitude to the proton amplitude 
between coherent and incoherent $\eta$-photoproduction on the deuteron 
due to a nonvanishing complex and energy dependent phase relation. 
\end{abstract}


\section{Introduction}
The photoproduction of $\eta$-mesons on a nucleon 
is an extremely interesting process 
because the $\eta$, being an isoscalar particle, can act as a hadronic 
isospin filter, i.e., only isospin $t=1/2$ resonances can couple
to an $\eta{}N$ state. Consequently, there exists no $\eta{}N\Delta$ vertex,
and any contribution of the $\Delta(1232)$ resonance, dominant in pion 
photoproduction, to 
$\eta$-mesonic processes is strongly suppressed. Thus, $\eta$-meson production 
is an important tool to study the rather small contributions of those  
$t=1/2$ resonances which lie above the $\Delta(1232)$ and which usually are 
overshadowed in other reactions like, e.g., in pion photoproduction
by the $\Delta$ resonance.

Moreover, the $\eta$-meson selects from the set of $t=1/2$ nucleon resonances
only the $S_{11}(1535)$, which has almost equal 
partial decay widths into the $\pi{}N$ and $\eta{}N$ channels, 
while all other resonances in this energy region decay predominantly into 
pionic channels.
 This property appears very peculiar in comparison to 
the slightly heavier $S_{11}(1650)$ resonance, which carries the same 
quantum numbers as the $S_{11}(1535)$, but does not couple to the 
$\eta{}N$ state at all. 
Thus experimentally one can exploit this property of the
$\eta$-meson to discriminate this particular resonance from the other 
$N^\ast$ resonances by simply selecting the $\eta{}N$ final state. 
This means, that $\eta$-photoproduction is specifically suited in order
to study the electromagnetic (e.m.) properties of the $S_{11}(1535)$ 
resonance.

The corresponding process on the deuteron is of considerable interest, 
because one hopes to obtain information about the unknown reaction on the 
neutron, considering the deuteron as an approximate neutron target in view 
of its weak binding. In order to extract this information, the incoherent 
process appears to be very suited, since in this case the reaction is 
dominated by the quasifree contribution, for which interference effects 
between the elementary amplitudes of proton and neutron are very small, 
so that the contributions from proton and neutron add incoherently 
to a very good approximation. On the other hand, the coherent process 
offers a special bonus, because the deuteron constitutes an e.m.\ isospin 
filter, which means that in the coherent reaction one selects the 
isospin $t=0$ channels, in other words, only the isoscalar excitation
strength determines the reaction. Thus, the coherent process will provide
information on this small quantity almost independently from the incoherent 
reaction, which clearly is dominated by the isovector amplitude. 
Moreover, whereas one obtains from the incoherent reaction
the moduli of the amplitudes only, i.e., their relative phases 
remain unknown, the coherent reaction allows to extract new information 
on these relative phases of the elementary amplitudes. This is of particular 
interest with respect to the question whether there exists a 
discrepancy between the coherent and incoherent photoproduction process as 
has been reported in \cite{HoffR}. Analyzing the experimental results 
by a fit to the two sets of data within the impulse approximation (IA), 
these authors found for the ratio of the isoscalar 
amplitude $A_s$ to the proton amplitude $A_p$ for the e.m.\ $\eta$-production
from the coherent data a value which was about a factor 2 larger
than the one extracted from the incoherent reaction, i.e., 
\beq
 (A_s/A_p)_{coh.} = 0.2 \qquad\mbox{and}\qquad
 (A_s/A_p)_{inc.} = 0.09\,. 
\eq
This discrepancy was one of our motivations for studying this reaction, 
and it will turn out, that 
the seeming inconsistency is a result of an oversimplified analysis 
of the coherent reaction. Obtaining information on the neutron amplitude 
from the reaction under consideration, however, is possible only 
if competing two-body contributions from rescattering and meson exchange 
currents are reliably known. It is the aim of the present work, to study 
such effects in greater detail. 

Two-body mechanisms have been neglected in a previous study \cite{EB}, 
restricting to the IA alone. 
In this work, various ingredients of the IA have been studied 
like, e.g., different choices 
of the neutron resonance amplitude and different prescriptions for the 
assignment of the invariant mass of the elementary amplitude. Rescattering 
mechanisms have first been considered in first order by Hoshi et 
al. \cite{Hoshi} who found large contributions explaining at least 
qualitatively the large experimental data of Anderson and 
Prepost \cite{AnP69}. In the mean time these experimental results have not 
been confirmed in more recent refined experiments \cite{HoffR}, where indeed 
much smaller cross sections have been found.
The large rescattering contributions in~\cite{Hoshi} came predominantly 
from $\eta$-exchange whereas $\pi$-exchange gave a very tiny contribution 
only. However, the quality of the approximations used in this work is 
difficult to assess. On the other hand, 
Halderson and Rosenthal \cite{HR} found later a 
much smaller rescattering effect within the one-loop approximation leaving 
the experimental results of \cite{AnP69} as a puzzle. But 
they confirmed the dominance of $\eta$-exchange over $\pi$-exchange. 
A better treatment of rescattering effects beyond the one-loop approximation 
within the multiple scattering approach of Kerman, McManus and Thaler 
has been reported by Kamalov {\em et al.} \cite{LT}, finding very small
two-body effects. However, based on the 
result of \cite{HR} they again have restricted the rescattering to 
$\eta$-exchange only, leaving out completely $\pi$-exchange, which 
in the present work we found to give an important contribution. In fact, the
relative importance of $\pi$- versus $\eta$-exchange is model 
dependent with respect to different choices of coupling 
strengths \cite{Fixprivate}. 

Therefore, we would like to stress the point that for a consistent description 
it is necessary to generate the two-body operators by the same elementary 
vertices, which determine the one-body contribution. Otherwise, defining 
the rescattering mechanisms independently, one loses any predicting power.
Previously, we had analyzed in \cite{RA} the two-body mechanisms for
the coherent photoproduction on the deuteron with purely static
nucleon-resonance interactions for which we found a sizeable reduction of 
the total cross section. In anticipation of the main result of the present 
work, we found that the introduction of retarded, and thus more appropriate 
interaction mechanisms leads to quite different effects. In addition to the 
question of the size of hadronic rescattering we also have investigated the 
role of two-body meson exchange currents (MEC) which have not been studied 
previously.

This work is structured as follows: In Sect.~\ref{elem} we will briefly 
sketch the elementary model for $\eta$-photoproduction on the nucleon, which 
we have taken essentially from \cite{BT}. In Sect.~\ref{deut} we will then 
incorporate this model into the two-nucleon system. In particular, we will 
discuss the two-body mechanisms arising from hadronic rescattering and from 
MEC\@. In Sect.~\ref{observables} we recall the definition of the observables 
of coherent meson photoproduction. The results are presented and discussed
in Sect.~\ref{discuss}. Finally we give in Sect.~\ref{zusf} a short summary 
and an outlook.

\section{Elementary Process}
\label{elem}

As first important step in the present work we will fix the elementary
photoproduction amplitude. But we will not use a simple effective Lagrangian
approach like in \cite{Mukho1}, where the undetermined parameters are fixed by
fitting directly the experimental $\eta$-production data.  Instead we have
taken the somewhat more ambitious coupled channel model of Bennhold and
Tanabe \cite{BT} in which the following open one-meson channels are
considered, i.e., the hadronic processes $\pi{}N\rightarrow\pi{}N$,
$\pi{}N\rightarrow\eta{}N$, and their e.m.\ analoga
$\gamma{}N\rightarrow\pi{}N$, and $\gamma{}N\rightarrow\eta{}N$.  Of course
this model predicts also the processes $\eta{}N\rightarrow\eta{}N$ and
$\eta{}N\rightarrow\pi{}N$, but these are not amenable to experimental
observations due to the lack of $\eta$-meson beams. Further processes yet to
be considered are the pion- and photo-induced two-pion production
$\pi{}N\rightarrow\pi\pi{}N$ and $\gamma{}N\rightarrow\pi\pi{}N$,
respectively. The presence of these reaction channels is treated in a
phenomenological manner only by assigning the resonances an effective
two-pion width as is discussed below (see Eq.~(\ref{twopiwidth})).  There
exists another dynamical calculation by Green and Wycech \cite{GuW} using the
$K$-matrix method. However, we prefer to use the model of~\cite{BT} since 
it allows in a natural way the continuation into the off-shell region as is 
required if one wants to incorporate the elementary amplitude into nuclei. 
This off-shell continuation is not always easy and well defined in a 
$K$-matrix approach. 

Such an involved analysis is in principle unavoidable, because of the 
appearance of the same resonances in the various reaction channels and the 
fact that each resonance possesses a hadronic width related by unitarity to 
the hadronic reactions. 
The model of Bennhold and Tanabe, being a separable resonance model, 
is still a simple effective one because of the limitation to
only pure resonant states or meson-nucleon states in the hadronic 
sector. No meson-resonance and multi-meson-nucleon configurations are allowed. 
While the nucleon 
is treated as a physical particle, the resonances are considered as bare 
ones, being dressed by loops from the open decay channels. Thus the 
$T$-matrices for the two hadronic 
reaction channels are given in the form
\beqa\label{tmatrizen}
T_{\pi{}N\rightarrow\pi{}N} 
    &=& \sum_R v^\dag_{\pi{}NR}\stackrel{\bullet}{g}_{R}v_{\pi{}NR}\,
 \quad\mbox{and}\quad T_{\pi{}N\rightarrow\eta{}N} 
    = \sum_R v^\dag_{\eta{}NR}\stackrel{\bullet}{g}_{R}v_{\pi{}NR}\,,
\eqa
where we include three resonances, namely 
$R\in\{P_{11}(1440)\}, D_{13}(1520), S_{11}(1535)\}$. These resonances
are the isospin $t=1/2$ nucleon resonances below and just above 
the $\eta$-meson production threshold, and are thus the most likely ones 
to affect 
the $\eta$-photoproduction reaction. The operators $v^\dag_{BNR}$ and
$v_{BNR}$ denote the meson emission and absorption 
vertices, respectively, as obtained from the following Lagrangians 
(see e.g. \cite{Mukho1})
\beqa
\label{lbns11}
{\cal L}_{B{}NS_{11}}&=&-ig_{B{}NS_{11}}\bar{\Psi}\Phi\Psi_{S_{11}} + h.c.,\\
\label{lbnp11}
{\cal L}_{B{}NP_{11}}&=&-ig_{B{}NP_{11}}\bar{\Psi}\gamma_5\Phi\Psi_{P_{11}}
  +  h.c., \\
\label{lbnd13}
{\cal L}_{B{}ND_{13}} &=& \frac{g_{B{}ND_{13}}}{m_\pi}
\bar{\Psi}^\nu_{D_{13}}\gamma_5\partial_\nu\Phi\Psi + h.c., 
\eqa
where $B\in\{\pi,\eta\}$, and $\Psi$ and $\Psi_R$ denote nucleon and 
resonance spinors, respectively. The bare resonance masses and the relation 
of the coupling constants to the ones of \cite{BT} are given in 
Table~\ref{BTfacs}. Detailed expressions are listed in 
Appendix~\ref{vertices}. 

\begin{figure}
\centerline{%
\epsfxsize=15.0cm
\epsffile{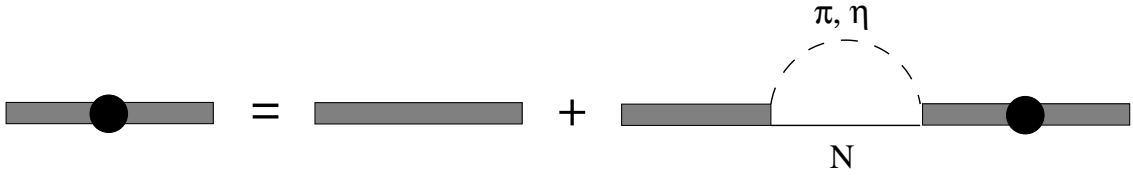}}
\vspace*{.3cm}
\caption{\label{dressedprop}
Dressing of a resonance propagator by $\pi$- and $\eta$-loops.} 
\end{figure}

The symbol $\stackrel{\bullet}{g}_{R}$ in (\ref{tmatrizen}) 
denotes a dressed resonance propagator containing in principle an infinite 
number of $\pi{}N$, $\eta{}N$ and $\pi\pi{}N$ loops. It is given as a 
function of the invariant energy $W$ by
\beqa
 \stackrel{\bullet}{g}_{R}(W)&=&\Big(W-m_R^0-\Sigma_R(W)+i\epsilon\Big)^{-1}\\
 &=& \Big(W-m_R(W) +\frac{i}{2}\Gamma_R(W)\Big)^{-1}\,, 
\eqa
where the energy dependent resonance mass $m_R(W)$ and the pionic and
$\eta$-mesonic parts of the resonance width $\Gamma_R(W)$ 
are related to real and imaginary parts of the resonance self energy 
$\Sigma_R(W)$, which arise from the above mentioned loop contributions. 
While the one-meson loops are evaluated explicitly within the present model 
(see Fig.~\ref{dressedprop}), 
the two-pion contributions are treated effectively only by parametrizing 
their imaginary part and incorporating the real part as constant 
in the bare mass $m_R^0$. Thus we have
\beqa
 m_R(W) &=& m_R^0 + \Re e\,\Sigma_R(W) \nonumber\\      
   &=& m_R^0 +\sum_{B\in\{\pi,\eta\}}
     {\cal P}\int\limits_{0}^{\infty}\!\frac{dq\,q^2}{(2\pi)^3}
     \frac{m_N}{2\omega_B(q)e_N(q)}\left(\frac{q}{m_B}\right)^{2L} 
     \frac{g^{[BT]\ 2}_{BNR}f^{[BT]\ 2}_{BNR}(q)}{W-\omega_B(q)-e_N(q)}\,,\\
 \Gamma_R(W)&=& -2\Im m\,\Sigma_R(W) \nonumber\\ 
   &=& \sum_{B\in\{\pi,\eta\}}
 \frac{1}{8\pi^2}
  \frac{m_N}{W} q_0 \left(\frac{q_0}{m_B}\right)^{2L}
    g^{[BT]\ 2}_{BNR} f^{[BT]\ 2}_{BNR}(q_0)
    +\Gamma^{\pi\pi}_R(W)\,, \label{BTwidth} 
\eqa
where $q_0=q_0(W)$ 
is the asymptotic meson momentum in the meson-nucleon
c.m.\ frame, $\omega_B(q)=\sqrt{m_B^2+q^2}$ is the on-shell energy of the
meson $B$, analogously $e_N(q)=\sqrt{m_N^2+q^2}$ is the energy of the nucleon,
and $L$ is the internal angular momentum of the resonance. 
Furthermore, 
$f^{[BT]}_{BNR}(q)$ denotes a hadronic form factor, which takes into account
effectively the internal structure of the baryons. Its functional form
\beqa
 f^{[BT]}_{BNR}(q) &=& 
     \Big[1+\left(\frac{q}{\Lambda_{BNR}}\right)^2\Big]^{-(1+L/2)}
\eqa
is chosen such, that the convergence of the loop integral is guaranteed.
For the two-pion contribution to the width we have adopted the effective 
treatment of \cite{BT,TBK94} and use a simple parametrization of the 
form
\beqa
 \label{twopiwidth}
 \Gamma_R^{\pi\pi}(W)&=&\gamma^{\pi\pi}_R\;\frac{W-m_N-2m_\pi}{m_\pi}\;
  \theta(\frac{W-m_N-2m_\pi}{m_\pi})\,.
\eqa

\begin{figure}
\centerline{%
\epsfxsize=9.0cm
\epsffile{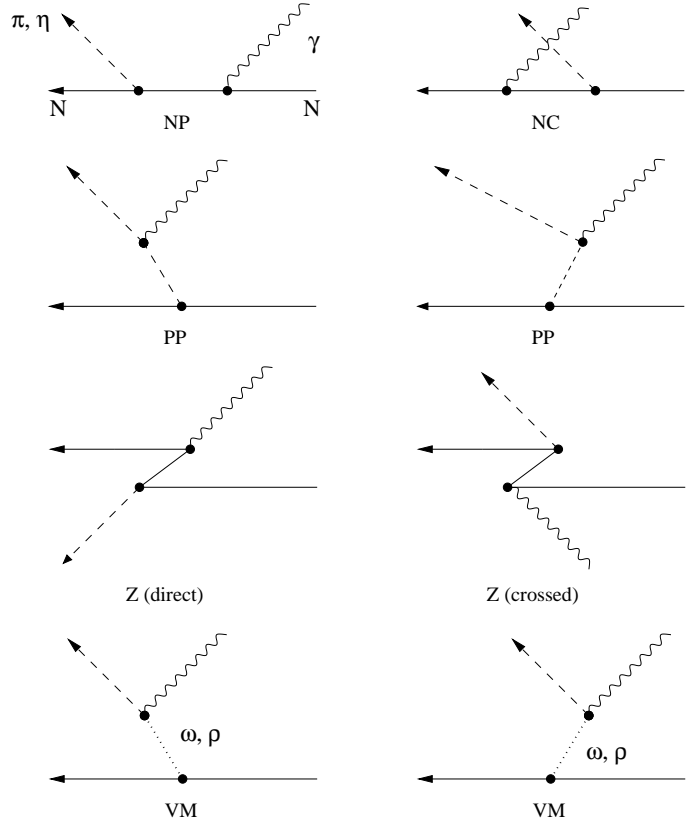}}
\vspace*{.3cm}
\caption{\label{overviewborn}
Diagrams of the contributions to the elementary Born amplitude: nucleon pole 
graph (NP), crossed nucleon pole graph (NC), pion pole graphs (PP), 
Z-graphs (Z), and vector meson exchange (VM).}
\end{figure}

\begin{figure}
\centerline{%
\epsfxsize=15.0cm
\epsffile{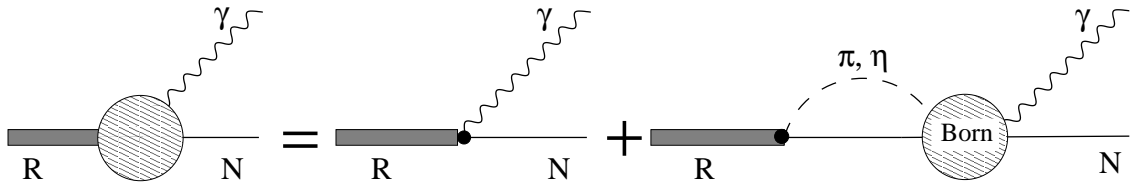}}
\caption{\label{elemresc}
Dressing of the e.m.\ resonance vertex by rescattering.} 
\end{figure}

\begin{figure}
\centerline{%
\epsfxsize=8.0cm
\epsffile{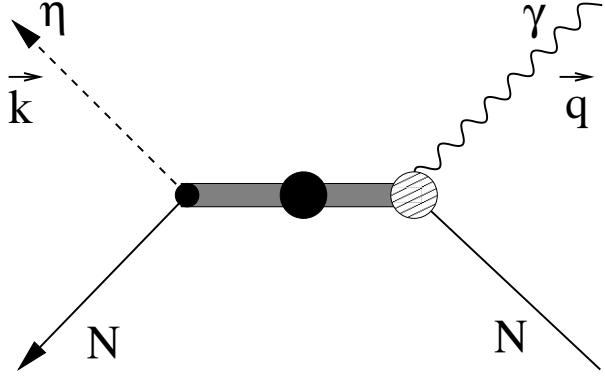}}
\vspace*{.3cm}
\caption{\label{elemreso}
Resonance contribution to the elementary meson photoproduction process 
including the dressed $\gamma NR$ vertex (see Fig.~\protect\ref{elemresc}).} 
\end{figure}

The elementary $\eta$-photoproduction amplitude is driven by a background 
from the Born terms and by a bare resonance excitation term describing 
$\eta$-photoproduction via intermediate bare resonance excitation. The Born 
contributions considered in this work are shown diagrammatically in 
Fig.~\ref{overviewborn}. The parameters of the Born terms are the same as 
used in~\cite{RA}, i.e., $g^2_{\eta N}/4\pi=0.4$ and the vector meson 
couplings from~\cite{EB,GK1}. 
The bare e.m.\ vertices $v_{\gamma NR}$ for 
resonance excitations are derived from the following Lagrangians
\beqa
\label{lgns11}
{\cal L}_{\gamma{}NS_{11}} &=& \frac{\hat{k}_{S_{11}}}{2(m_N+m^0_{S_{11}})}
\bar{\Psi}_{S_{11}}\gamma_5\sigma_{\mu\nu}\Psi\,F^{\mu\nu} + h.c., \\
\label{lgnp11}
{\cal L}_{\gamma{}NP_{11}} &=& 
 -\frac{\hat{k}_{P_{11}}}{2 m^0_{P_{11}}}
 \bar{\Psi}_{P_{11}}\sigma_{\mu\nu} \Psi F^{\mu\nu} + h.c., \\
\label{lgnd13}
{\cal L}_{\gamma{}ND_{13}} &=& \frac{i \hat{k}_{D_{13}}}{2m_N}
\bar{\Psi}_{D_{13},\nu}\gamma_\mu\Psi\,F^{\mu\nu} + h.c.,
\eqa
where $F^{\mu\nu}$ denotes the e.m.\ field tensor. Furthermore, the e.m.\ 
couplings $\hat{k}_R$ contain isoscalar and isovector contributions
\beq
 \hat{k}_R = g^{(0)}_{\gamma{}NR}{\mathbb 1} + g^{(1)}_{\gamma{}NR}\tau_0.
\eq
In a dynamical treatment, the bare e.m.\ vertices become dressed
by hadronic rescattering as is illustrated in Fig.~\ref{elemresc}, i.e.,
$v_{\gamma NR} \rightarrow \widetilde v_{\gamma NR}(W)$ \cite{Ohta1,WiW98}. 
The dressing of the e.m.\ vertices leads to complex, energy dependent 
couplings. This fact follows directly from the 
unitarity relation demanding such loop diagrams. 
Thus the total photoproduction amplitude reads
\beqa\label{gammatmatrizen}
T_{\gamma{}N\rightarrow\eta{}N} 
    &=& T_{\gamma{}N\rightarrow\eta{}N}^{B} +
\sum_R v^\dag_{\eta{}NR}\stackrel{\bullet}{g}_{R}\widetilde v_{\gamma{}NR}\,,
\eqa
where $T_{\gamma{}N\rightarrow\eta{}N}^{B}$ denotes the Born contribution.
The resonance part is shown diagrammatically in Fig.~\ref{elemreso}.

\begin{figure}
\centerline{%
\epsfxsize=8.0cm
\epsffile{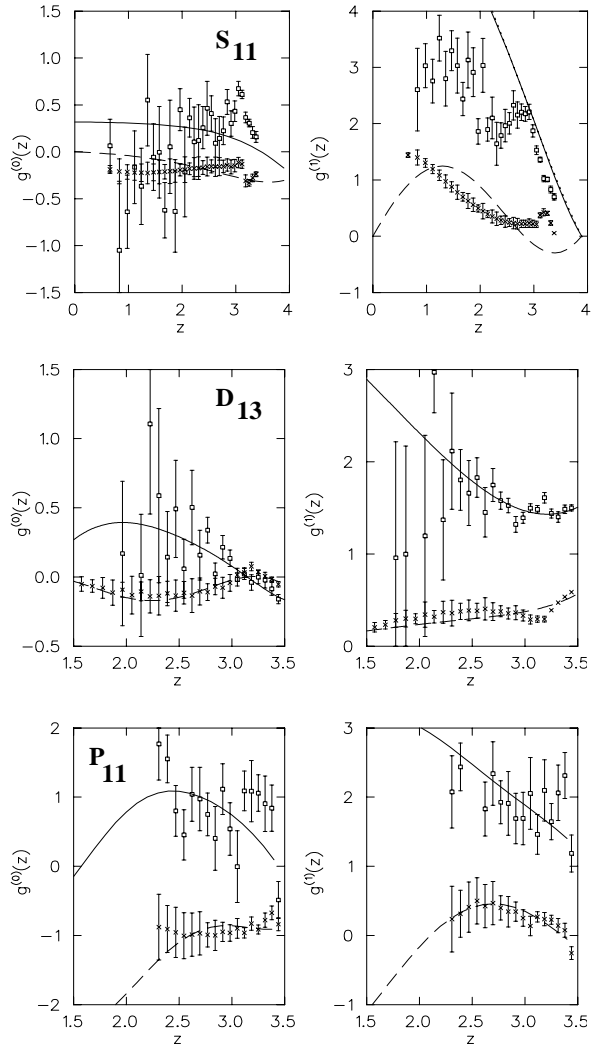}}
\vspace*{.3cm}
\caption{\label{effcoupl} Effective e.m.\ coupling strength of the
nucleon resonances extracted from the experimental multipoles 
{\protect\cite{SAID}}. Notation: $\Box$ real part of the effective
couplings; $\times$ imaginary part of the effective couplings;
full curves: fit of the parametrization of Eq.~(\protect\ref{fitfuncs}) to 
the real part; dashed curves: fit to the imaginary part.}
\end{figure}

In this work we do not calculate these 
loop contributions explicitly, but follow Bennhold and Tanabe by fitting 
each pion photoproduction multipole $X$, to which a given resonance 
$R$ contributes, to its experimental value $X_{exp}$ from which the 
Born contribution $X_{Born}$ has been subtracted, by defining an effective
e.m.\ coupling  
\beq
 g_{\gamma{}NR}^{(I)}=\frac{X_{exp}-X_{Born}}{X_R(g_{\gamma{}NR}^{(I)}
 \rightarrow 1)},
\eq
where $X_R(g_{\gamma{}NR}^{(I)}\rightarrow 1)$ is the 
purely resonant multipole with the e.m.\ coupling set equal to one. 
For the fit we use the following complex parametrization 
\beqa
\label{fitfuncs}
  g^{(I)}_{\gamma{}NR}(W)&=& |g^{(I)}_{\gamma{}NR}(W)|\,%
\exp(i \Phi^{(I)}_{\gamma{}NR}(W)),
\eqa
where modulus and phase are described by polynomials in $z=k_\pi(W)/m_\pi$
\beqa
\label{para1}
  |g^{(I)}_{\gamma{}NR}(W)|  &=& a^{(I)}+b^{(I)}\,z+c^{(I)}\,z^2
    + d^{(I)}\,z^3+e^{(I)}\,z^4, \\
\label{para2}
  \Phi^{(I)}_{\gamma{}NR}(W)  &=& z\,\left(f^{(I)}+g^{(I)}\,z
    + h^{(I)}\,z^2\right)\,,
\eqa
and $I=0,1$ denotes isoscalar and isovector excitations, respectively.
The open parameters are fit to the elementary photoproduction data, i.e., 
the pion photoproduction multipoles $E^{(0),(1/2)}_{0+}$, $M^{(0),(1/2)}_{1-}$,
$E^{(0),(1/2)}_{2-}$, and the total cross section of $\eta$-production on
the proton. The results of the fit are
shown in Fig.~\ref{effcoupl} and the corresponding parameters of 
Eqs.~(\ref{para1}) and (\ref{para2}) are summarized in Table~\ref{gefitte}.
The fit certainly is not of high precision, which is not the aim of the 
present work, but it is of sufficient quality (see the discussion of 
observables below) for our purpose, namely to assess the relative importance 
of interaction effects. With respect to our previous work~\cite{RA} we would
like to remark, that the present fit differs from the one in~\cite{RA} 
because there the Born amplitude contained a small error resulting in a 
slightly different fit with different parameters. But the description of the 
observables of the elementary process is of the same quality. Also the size of
interaction effects was not affected by this error. 

With respect to unitarity we must state that our model, and also the original
work of Bennhold and Tanabe, is not unitary, although the hadronic resonance 
model is per construction two-body unitary below the two-pion threshold.
The effective treatment of the two-pion channel and the parametrization 
of the dressed e.m.\ vertices instead of evaluating the dressing loops 
destroys unitarity. In order to fulfil unitarity one would need to include a 
dynamical description of e.m.\ two-pion production and its hadronic analogon,
pion-induced two-pion production. Such a dynamical treatment of 
two-pion production is quite involved. 
For this reason, to our knowledge, there does not exist any calculation of 
meson production in this energy region fulfilling unitarity. 
In view of the fact that our main emphasis lies on the $\eta$-photoproduction 
on the deuteron, we believe that the present effective description is 
justified. 

Another remark is in order with respect to the interpretation of the 
parameters of effective models in view of the fact that 
there exists quite a number of different models in the literature.
One should be extremely cautious in the interpretation of the resonance 
parameters in terms of microscopic nucleon resonance models because they are 
in general model dependent quantities, and thus are not observable.
None of the effective models available today
offers the possibility to extract resonance parameters in a model independent
way. The reason for this is an inherent unitary ambiguity of such
approaches, which makes it impossible to separate uniquely background and 
resonant contributions (see Wilhelm {\em et al.} \cite{WiW96}).

\begin{figure}
\centerline{%
\epsfxsize=8.0cm
\epsffile{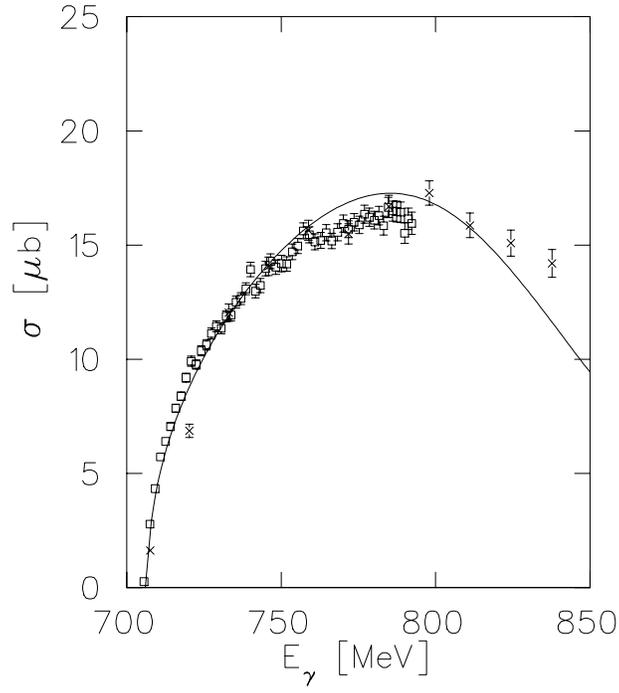}}
\vspace*{.3cm}
\caption{\label{elem1} Total cross section of 
$\eta$-photoproduction on the proton. Experimental data: 
$\Box$: Krusche {\em et al.}~{\protect\cite{Kru95}},
$\times$: Wilhelm {\em et al.}~{\protect\cite{Wilex}}.}
\end{figure}

\begin{figure}
\centerline{%
\epsfxsize=9.0cm
\epsffile{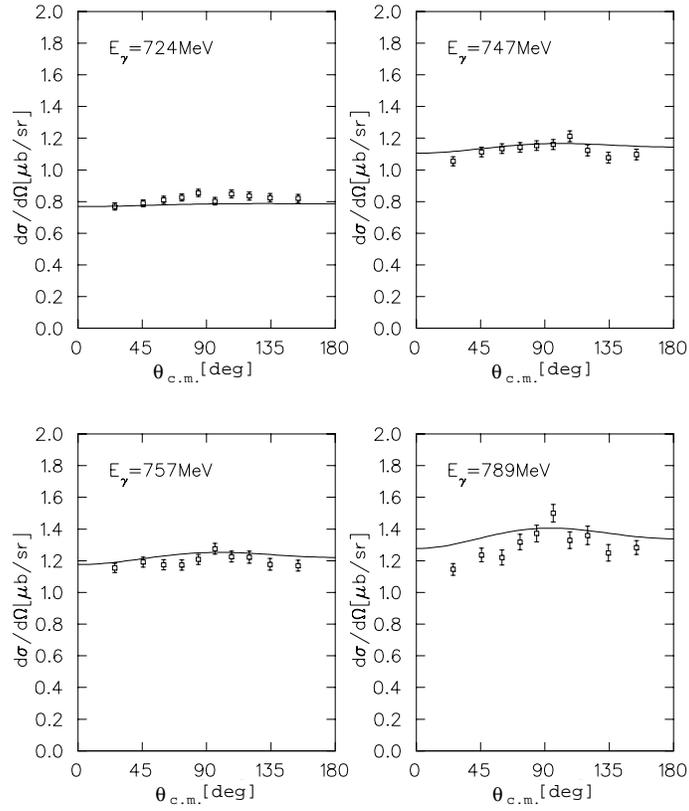}}
\vspace*{.3cm}
\caption{\label{elem2}
Differential cross section of $\eta$-photoproduction on the proton.
Experimental data from~\protect\cite{Kru95}}
\end{figure}

The quality of the description of the data of the elementary process by our 
model can be seen in Figs.~\ref{elem1} and \ref{elem2}. 
It is quite good for the total cross section of 
$\eta$-photoproduction on the proton (Fig.~\ref{elem1}). Only above a photon 
energy of about 750 MeV one notices a slight overestimation of the data 
from \cite{Kru95}. Similarly, the angular dependence of the unpolarized 
differential cross section in Fig.~\ref{elem2} is described quite 
satisfactorily. The theory shows a slightly more isotropic behaviour than 
the data, and at the highest energy a small overall shift to higher values 
corresponding to the slight overestimation of the total cross section. But
we do not consider this deviation as a serious defect which is also found in
other approaches, for example in \cite{LT}. 
One important result with respect to the question of the strength of the 
scalar amplitude is that we found for the proton and neutron amplitude at the
resonance energy in the present model the complex values
\beq
 A_n = ( -114-i1.7 )\times 10^{-3} \mbox{GeV}^{-1/2},
\quad  A_p =  ( 120.9-i66.1 )\times 10^{-3} \mbox{GeV}^{-1/2}, 
\eq
from which one obtains for the ratios of the total cross sections on neutron
and proton as well as for $A_s/A_p$ the values
\beq
 (\sigma_n/\sigma_p)_{res} = |A_n/A_p|^2 = 0.68 \approx 2/3,
 \quad A_s/A_p = 0.25\,e^{-i 0.969} \,,
\eq
respectively,
where the modulus of $A_s/A_p$ essentially agrees with the value extracted
from the coherent process within the impulse approximation
but the phase is different from 0 and $\pi$. 
The neglect of this nonvanishing phase in the analysis of \cite{HoffR} 
appears to be the origin of the above mentioned disagreement between the 
ratios extracted from the coherent and incoherent reactions.

\section{Process on the Deuteron}
\label{deut}
\begin{figure}
\centerline{%
\epsfxsize=9.0cm
\epsffile{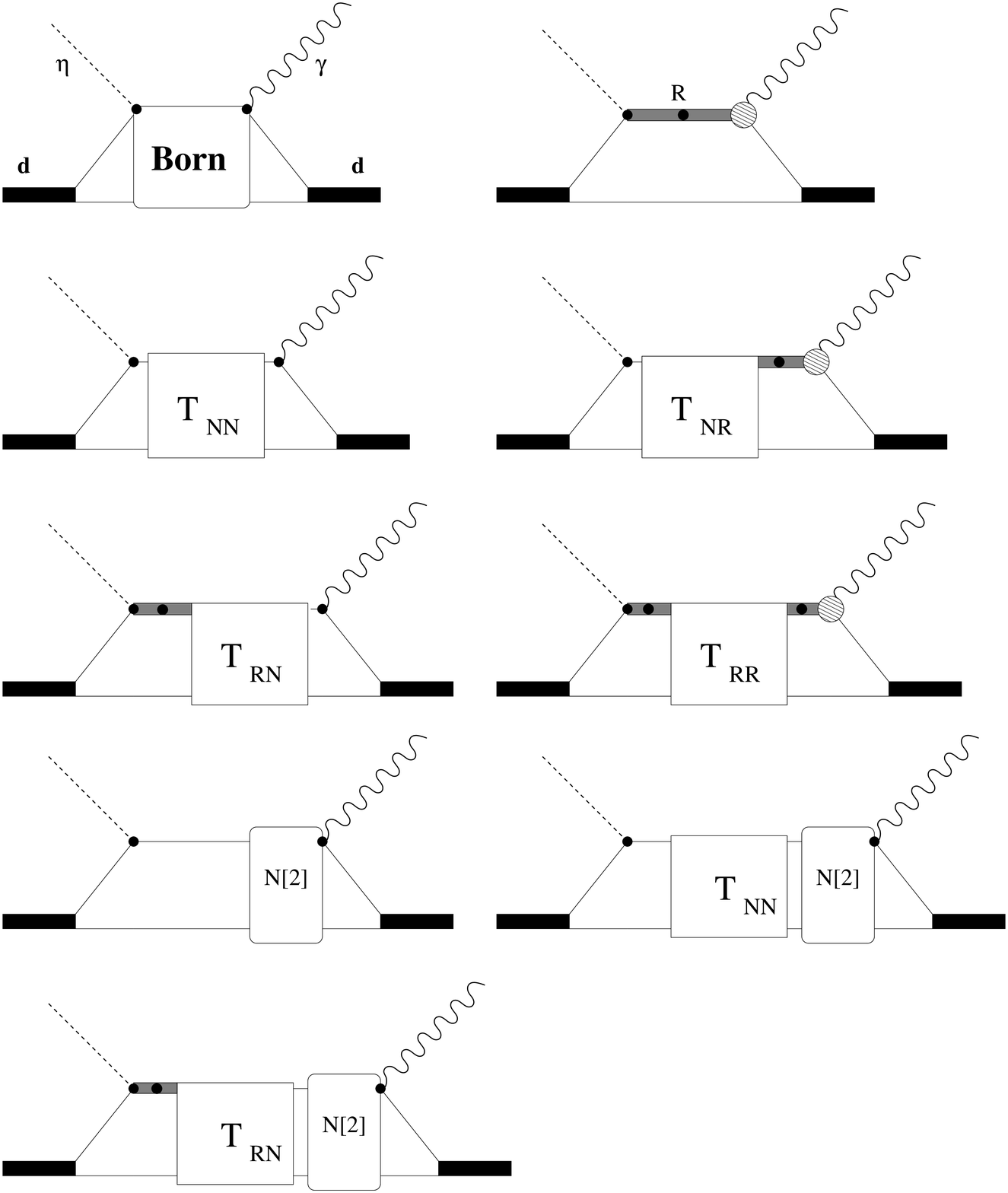}}
\vspace*{.3cm}
\caption{\label{overview}
Diagrammatical overview of the model of coherent $\eta$-photoproduction 
on the deuteron. 
The box labeled Born contains also disconnected diagrams where the photon is
absorbed by one nucleon and the $\eta$ is emitted by the other.
Hadronic rescattering is indicated by the square boxes 
labeled $T_{NN}$, $T_{NR}$, $T_{RN}$, and $T_{RR}$. Meson exchange current 
contributions are indicated by the boxes labeled $N[2]$.} 
\end{figure}

For the photoproduction on the deuteron we include in addition to the 
impulse approximation, i.e., the one-body contribution, 
various two-body diagrams which arise (i) from the off-shell behaviour
(disconnected Born diagrams), 
(ii) hadronic rescattering between photon absorption and meson emission, and
(iii) from two-body meson exchange currents. A diagrammatical overview 
of the various contributions considered in this work is given in 
Fig.~\ref{overview}. The first two diagrams describe the impulse 
approximation comprising the Born and resonance contributions, the former 
including the disconnected graphs and the latter containing 
the dressed photon vertex. The next four diagrams comprise the 
various hadronic interactions of the intermediate two-baryon states 
including nucleon-resonance transition interactions. The last three 
diagrams describe the MEC contributions combined with hadronic rescattering. 

For the impulse approximation we have to embed
the elementary photoproduction amplitude into the two-nucleon system. 
To this end we need this amplitude full off-shell in an arbitrary frame of 
reference. This can be achieved in our model by a straightforward construction
from the appropriate time ordered diagrams 
using the Lagrangians given in Eqs.~(\ref{lbns11})--(\ref{lbnd13}) and 
(\ref{lgns11})--(\ref{lgnd13}).
It is in contrast to other approaches, where the elementary amplitude 
is constructed first on-shell in the photon-nucleon c.m.\ frame with 
subsequent boost into an arbitrary reference frame and some prescription 
for the off-shell continuation. In the latter method, one loses terms which 
by chance vanish in the c.m.\ frame \cite{EB}. In our approach,
the only uncertainty arises from the assignment of the invariant energy
for the photon-nucleon subsystem in the resonance propagators as has 
been discussed in detail in \cite{EB}. Here we use the spectator 
on-shell approach as in \cite{PW1}. 

As already mentioned, the Born currents are constructed from the 
off-shell expressions of the corresponding elementary operators. 
The construction is straightforward and explicit formulae can be 
found in \cite{Ritz}. A remark is in order with respect to the vector 
meson contribution. The expressions in \cite{Ritz} differ from those 
in \cite{EB}, where the vector meson contribution was derived from 
on-shell Feynman diagrams with implicit time ordering. Because in the 
present process both nucleon lines are off-shell, this method is, 
strictly speaking, not applicable. However, in view of the very small energy 
transfer of the vector meson, this approximation
turns out to be quite reliable.

The hadronic rescattering mechanisms are treated by solving a system of
coupled Lippmann-Schwinger equations in the space of $NN$ and various 
isobar configurations ($NR$) neglecting $RR$ configurations, i.e. 
\beq
 T = V + V\,G_0\,T\,,
\eq
where $T$-matrix, potential $V$ and free propagator $G_0$ are matrices with 
respect to the various two-baryon channels
\beqa
\label{LGS}
T&=&
\left(\begin{array}{cccc}
 T_{NN\leftarrow NN}   & T_{NN\leftarrow NR_1}   & \dots & T_{NN\leftarrow NR_n} \\
 T_{NR_1\leftarrow NN} & T_{NR_1\leftarrow NR_1} & \dots & T_{NR_1\leftarrow
 NR_n} \\
 \vdots & & \ddots & \vdots \\
 T_{NR_n\leftarrow NN} & T_{NR_n\leftarrow NR_1} & \dots & T_{NR_n\leftarrow
 NR_n}
\end{array}\right), 
\eqa
a corresponding matrix for the potential $V$, and 
\beqa
G_0 &=&
\left(\begin{array}{cccc}
 G_{0}^{NN} & 0  & \dots & 0 \\
  0 & \stackrel{\bullet}{G}^{NR_1}_{0} & \dots & 0 \\
 \vdots & & \ddots & \vdots \\
 0 & 0 & \dots & \stackrel{\bullet}{G}^{NR_n}_{0} \label{G0}
\end{array}\right).
\eqa
Here, $\stackrel{\bullet}{G}^{NR}_{0}$ denotes the $NR$ propagator in the 
c.m.\ system with a dressed resonance 
\beq
 \stackrel{\bullet}{G}^{NR}_{0}(\vec p,W) 
 =\Big(W-m_N-\frac{\vec p^{\;2}}{2m_N}-m_R^0-\frac{\vec p^{\;2}}{2m_R^0}
      - \Sigma_R(W_{sub}) 
      +i\epsilon\Big)^{-1},
\eq
where $\vec{p}$ denotes the relative two-baryon momentum. 

\begin{figure}
\centerline{%
\epsfxsize=9.0cm
\epsffile{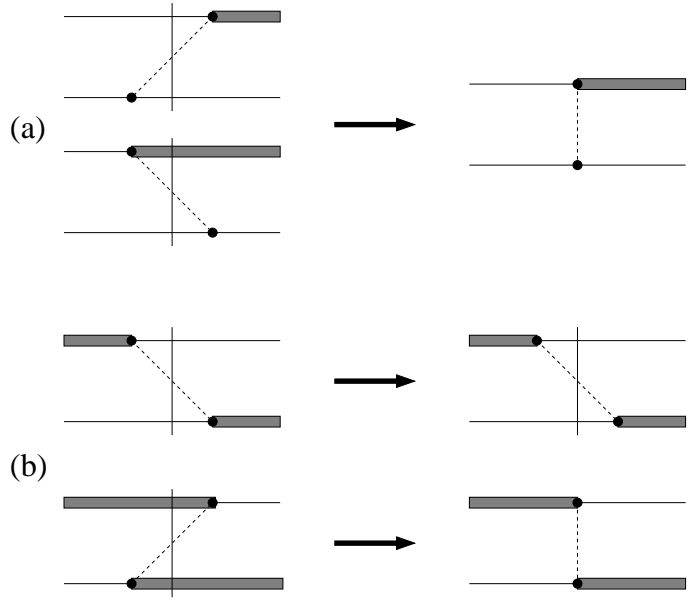}}
\vspace*{.3cm}
\caption{\label{timeorders} Treatment of the different time orderings of the 
hadronic transition potentials. 
(a) static approximation for the $NN\leftrightarrow{}NR$ potential,
(b) upper part: retarded $NR$ exchange potential,
 lower part: static approximation of the meson-RR propagator.}
\end{figure}

For the $NN$ potential we take a realistic potential which has to be 
renormalized as outlined below. 
The $NR$ transition ($V_{NR\leftarrow{}NN}$) and diagonal 
($V_{RN\leftarrow{}NR}$) potentials are
constructed from the usual time ordered diagrams (see Fig.~\ref{timeorders}) 
using the elementary vertices from the Lagrangians (\ref{lbns11}) through 
(\ref{lbnd13}). As diagonal interaction we take the exchange contribution 
only, where nucleon and resonance are interchanged, and neglect the 
non-exchange part in view of unknown coupling strengths. Thus the potentials 
have the general form
\beqa
 V_{NR\leftarrow{}NN} &=& \sum_{B\in\{\pi,\eta\}}
 {\Omega}_B \Omega_{q}^{NR;NN}(1,2) 
 \left( G_0^{BNN}(W) + G_0^{BNR}(W) \right)
  + (1\leftrightarrow{}2),\\
 V_{RN\leftarrow{}NR} &=& \sum_{B\in\{\pi,\eta\}}
 {\Omega}_B \Omega_{q}^{RN;NR}(1,2) 
 \left( G_0^{BNN}(W) + G_0^{BRR}(W) \right)
  + (1\leftrightarrow{}2),
\eqa
where $R\in\{P_{11},D_{13},S_{11}\}$, 
$\Omega_{q}(1,2)$ denotes a momentum space operator depending
on the spin and momentum variables of the participating baryons,
and ${\Omega}_B$ is an isospin operator
\beq
\begin{array}{ll}
 {\Omega}_\pi = \vec{\tau}_1\!\cdot\!\vec{\tau}_2 
 & \mbox{\ for $\pi$-exchange}, \\
 {\Omega}_\eta = {1} & \mbox{\ for $\eta$-exchange}.
\end{array}
\eq
Furthermore, $G_0^{BNN}(W)$, $G_0^{BNR}(W)$, and $G_0^{BRR}(W)$ denote the 
meson-$NN$, meson-$NR$ and meson-$RR$ propagators, respectively, 
\beq
\label{meson-baryon-prop}
G_0^{BXY}(W)=\Big(W-e_X(\vec p^{\,\prime})-e_Y(\vec p\,)
-\omega_B(\vec p^{\,\prime}-\vec p\,)+i\epsilon\Big)^{-1},
\eq
where $(XY)\in \{(NN),(NR),(RR)\}$. 
The nonrelativistic on-shell energies of the baryons are defined as
$e_R(\vec p\,)=m^0_R + {\textstyle\frac{\vec p^{\,2}}{2m^0_R}}$ and
$e_N(\vec p\,)=m_N + {\textstyle\frac{\vec p^{\,2}}{2m_N}}$.
Explicit expressions of the potential operators 
are listed in Appendix~\ref{transpot}. For consistency, the coupling 
constants are taken from the Bennhold-Tanabe model as defined by the 
Lagrangians in Eqs.~(\ref{lbns11}) through (\ref{lbnd13}) and listed 
in Table~\ref{BTfacs}. 

The meson-$NN$-propagators are taken either in the static approximation, 
or fully retarded in order to study the role of meson retardation. 
On the other hand, we treat the meson-$NR$ and the 
meson-$RR$ propagators always in the static approximation including the 
mass differences of the participating baryons (see Fig.~\ref{timeorders}),
i.e.
\beqa
 G^{BXY}_0(W)
 &\rightarrow& \Big(2m_N-m^0_{X}-m^0_{Y}-\omega_k\Big)^{-1}\,.
\label{stat-meson-baryon-prop}
\eqa

\begin{figure}
\centerline{%
  \epsfxsize=8.0cm \epsffile{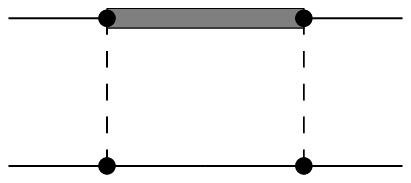}}
\vspace*{.3cm}
\caption{\label{nnren}
Static nucleon-resonance box used in the renormalization scheme of Green and
Sainio.}
\end{figure}

\begin{figure}
\centerline{%
  \epsfxsize=12.0cm \epsffile{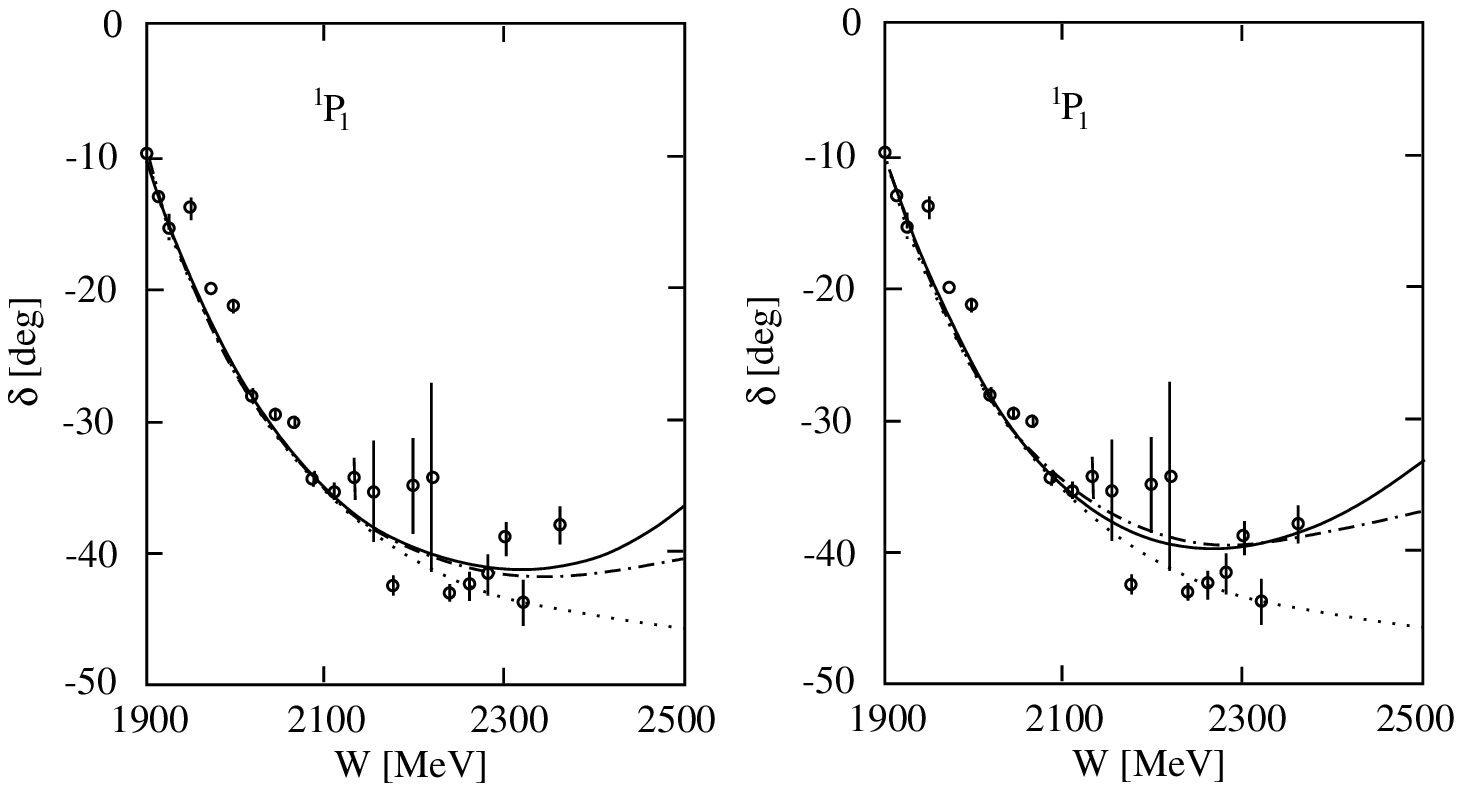}}
\vspace*{.3cm}
\caption{\label{p-wave-box}
$^1P_1$-phase shift of $NN$-scattering as function of the invariant 
energy $W$ of the $NN$-system. Notation of curves: dotted: OBEPQ-B 
$NN$-potential,
dash-dot: coupled channel with static rescattering and box renormalization, 
solid: coupled channel with retarded rescattering and box renormalization. 
Left panel: rescattering through $NS_{11}$ only, right panel: complete 
rescattering. The data are from the VPI analysis~\protect\cite{SAID}.}
\end{figure}

At the end of this section we will briefly describe the above 
mentioned renormalization of a realistic potential.
With the introduction of additional isobar configurations $NR$ with 
corresponding interactions into a coupled channel approach, one changes 
the effect of the
 interaction on the $NN$ channel, which was originally described 
by the $NN$-potential acting in the pure $NN$ space alone and which was 
fit to $NN$ scattering data and deuteron properties. Thus, the good 
agreement with experiment is destroyed. In order to avoid this feature, 
there are two possible
solutions. Either one could fit all parameters of the extended interaction
model, pure nucleonic as well as resonance parameters, to the $NN$ data. 
However, such a fit procedure is quite involved and time consuming, and is 
beyond our scope at the moment. Or one could ``renormalize'' the original 
$NN$-potential in such a way that together with the
additional interactions one reproduces the effect of the original potential. 
Such a renormalization recipe was introduced by Green and Sainio \cite{GS} by
subtracting a static $NR$ box at a fixed, appropriately chosen energy 
(see the diagram in Fig.~\ref{nnren}). In the present work such a box 
renormalization at the energy of $W=2m_N$ has been applied. 
However, it is obvious that the first method should be preferred in principle,
because the box renormalization method is approximate and valid over a 
limited energy range only. In order to demonstrate the quality of the box 
renormalization we show as one example in Fig.~\ref{p-wave-box} the phase 
shift for the $^1P_1$-partial wave, which is the most important partial wave 
for the rescattering contribution, because it is the only isoscalar partial 
wave, which couples to a $NS_{11}$-$S$-wave. One readily notices that 
in the energy range, where 
the original $NN$-potential has been fit, the good description is preserved. 
This is valid also for the other partial waves (for details see~\cite{Ritz}).

\section{Definition of Observables}
\label{observables}
Before we discuss the results of our calculation, we will 
give a short sketch of the definition of the observables of
$\eta$-photoproduction which we restrict to beam and target polarization, 
neglecting possible recoil polarization. The general form of an 
observable can be found in \cite{Are99}. 

\begin{figure}
\centerline{%
\epsfxsize=9.0cm
\epsffile{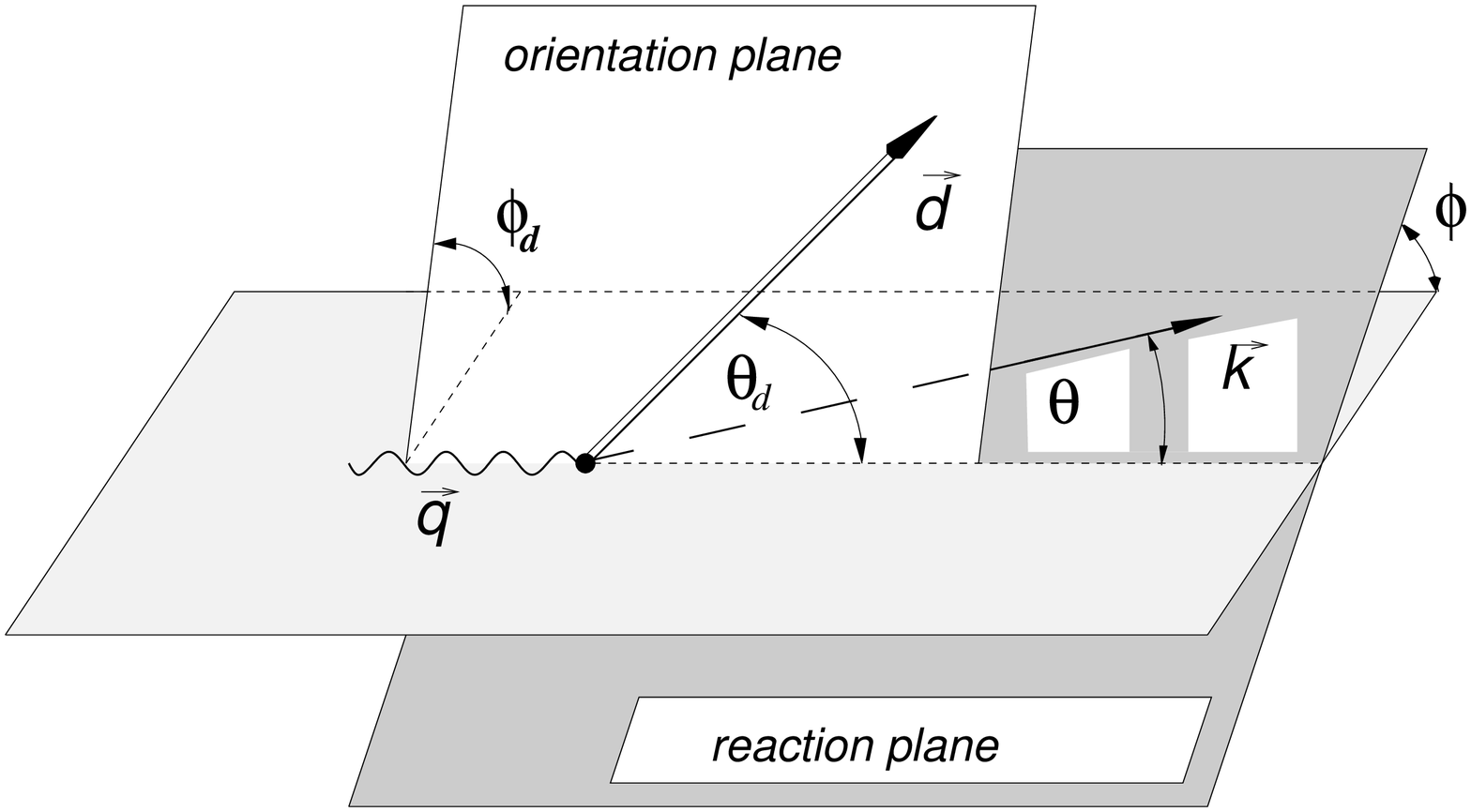}}
\vspace*{.3cm}
\caption{\label{becks}
Kinematical variables of the coherent
  $\eta$-photoproduction process on the deuteron.} 
\end{figure}

We choose our frame of reference with the $z$-axis pointing in the
direction of the photon momentum $\vec{k}$ which also serves as 
quantization axis for the deuteron spin states. 
The direction of the $x$-axis is defined by the density matrix of the 
photon polarization with respect to the basis of circular polarization states
\beqa
 \rho^\gamma_{\lambda\lambda^\prime} 
    = \frac{1}{2}\left(\delta_{\lambda\lambda^\prime} 
        +\vec{P}^\gamma\cdot\vec{\sigma}_{\lambda\lambda^\prime} \right),
\qquad \lambda,\lambda^\prime =\pm 1\,,
\eqa
where $\vec{\sigma}$ denotes the Pauli spin operator, and 
$\vec{P}^\gamma$ characterizes the polarization of the photon. 
In detail, $P^\gamma_c=P^\gamma_z$ describes the degree of circular 
polarization, while 
$P^\gamma_\ell=\sqrt{(P^\gamma_x)^2 + (P^\gamma_y)^2}$ describes 
the one of linear polarization. Now the $x$-axis is chosen in the direction 
of maximal linear polarization, i.e., $P^\gamma_x=-P^\gamma_\ell$ and 
$P^\gamma_y=0$. Furthermore, the direction of the outgoing meson momentum 
$\vec q$ is characterized by the angles $(\theta,\phi)$. It defines together 
with the photon momentum the reaction plane. The geometry is shown in 
Fig.~\ref{becks}. If the incoming photon beam is not linearly polarized, then
the $x$-axis may be chosen arbitrarily, as there is no dependence on
the angle $\phi$.

A possible target orientation is described by the following density matrix
\beqa
  \rho^d_{m^\prime m} &=& \langle 1\,m^\prime \mid \rho^d \mid 1\,m \rangle 
 \nonumber \\
 &=& \frac{1}{\sqrt{3}} (-)^{1-m^\prime} \sum_{I=0}^2\sum_{M=-I}^I 
 \hat{I} 
 \left(\begin{array}{lll} 1 & 1 &I \\ m & -m^\prime &M\end{array}\right) 
 P^{d\;\ast}_{IM},
\eqa
where the eight independent parameters $P^d_{IM}$ ($P^d_{00}=1$ by definition) 
describe the orientation of the target. 
For present experimental methods for deuteron orientation there exists an 
axis $\vec{d}$, characterized by angles $(\theta_d,\phi_d)$ with respect to 
which the density matrix is diagonal. The orientation axis $\vec{d}$ defines 
the orientation plane as also indicated in Fig.~\ref{becks}. Then, besides 
the orientation angles, 
one has only two independent parameters $P^d_1$ and $P^d_2$. They are 
related to the probabilities $p_{\pm{}1}$ to find the projections
$m_d=\pm{}1$ along the axis $\vec{d}$ by
\beqa
 P^d_1 &=& P^d_{10} = \sqrt{\frac{3}{2}}(p_1-p_{-1}), \\
 P^d_2 &=& P^d_{20} = \frac{1}{\sqrt{2}}\left(3(p_1+p_{-1})-2\right), 
\eqa
and one has
\beqa
 P^d_{IM} &=& P^d_I \mbox{e}^{iM\phi_d} d^{I}_{M0}(\theta_d)\,.
\eqa

Formal expressions for the differential cross section in coherent 
pseudoscalar meson photoproduction from an oriented deuteron target have 
been given in \cite{PW1,BBB95} in terms of beam, target and beam-target 
asymmetries $\Sigma$, $T_{IM}$, and $T^{c/l}_{IM}$, respectively. Here 
we follow the more general approach of \cite{Are99}. 
The general form of the differential cross section can be described by the 
unpolarized cross section and various asymmetries, which depend on the 
scattering angle $\theta$ only 
\beqa
 \frac{d\sigma}{d\Omega}  &=& \frac{d\sigma_0}{d\Omega} \,
\sum _{I=0}^2 P_I^d \sum _{M=0}^I
 \Big\{\Big(\widetilde T_{IM} +P^\gamma_\ell 
 \widetilde T^\ell_{IM+} \cos 2\phi\Big)
\cos (M\tilde{\phi}-\delta_{I1} {\pi \over 2}) \nonumber\\
& &\qquad\qquad+\Big(P^\gamma_c \widetilde T^c_{IM}+
P^\gamma_\ell \widetilde T^\ell_{IM-}\sin 2\phi\Big) 
\sin (M\tilde{\phi}-\delta_{I1} {\pi \over 2})
\Big\} d_{M0}^I(\theta_d)\,,\label{obsfin}
\eqa
where $\tilde{\phi}=\phi-\phi_d$. The unpolarized cross section and the 
asymmetries are defined by
\beqa
\frac{d\sigma_0}{d\Omega}&=&2\,{\cal U}^{11 00}_{00}\,,\\
\frac{d\sigma_0}{d\Omega} \,\widetilde T_{IM}&=&\frac{4}{1+\delta_{M0}}
\Re e\left(i^{\delta_{I1}}
{\cal U}^{11 I M}_{00}\right)\,,\\
\frac{d\sigma_0}{d\Omega} \,\widetilde T^c_{IM}&=&
 \frac{4}{1+\delta_{M0}}\,\Re e\left(i^{1+\delta_{I1}}
{\cal U}^{11 I M}_{00}\right)\,,\\
\frac{d\sigma_0}{d\Omega} \,\widetilde T^\ell_{IM\pm}&=&
\mp\frac{2}{1+\delta_{M0}}\,\Re e\left[i^{\delta_{I1}}
\left({\cal U}^{-11 I M}_{00}\pm(-)^{I+M}
{\cal U}^{-11 I -M}_{00}\right)\right]\,,
\eqa
with
\beqa
{\cal U}_{00}^{\lambda' \lambda I M}&=& 
\frac{c}{\sqrt{3}}\hat{I}\sum_{m^\prime m n}(-)^{1-m}
      \left(\begin{array}{ccc} 1 & 1 & I \\ 
            n & -m& M\end{array}\right) 
t^\ast_{m^\prime \lambda' n}\,t^{\vphantom{\ast}}_{m^\prime \lambda m}\,,
\label{UIM}
\eqa
and $c$ is a kinematical factor
\beq
 c = \frac{1}{16\pi^2}\;\frac{k}{q}\;
\frac{\sqrt{m_d^2+q^2}\sqrt{m_d^2+k^2}}{W^2}.
\eq
Note that $\widetilde T_{00}=1$, $\widetilde T^\ell_{I0-}=0$ for $I=0,2$, 
and $\widetilde T^\ell_{10+}=0$. 
In (\ref{UIM}), the ``small'' $t$-matrix elements 
are defined by separating the $\phi$-dependence from the $T$-matrix elements 
\beq
 T_{m^\prime \mu m}(\theta,\phi)
    = \mbox{e}^{i(\mu+m)\phi} \,t_{m^\prime \mu m}(\theta).
\eq
They have the following symmetry property 
\beq
  t_{-m^\prime -\mu -m}(\theta)
  =(-)^{1+m^\prime+\mu+m} \,t_{m^\prime \mu m}(\theta).
\eq
With respect to the asymmetries defined in \cite{PW1}, we note the following 
relations to the ones introduced above
\beqa
T_{IM} &=& (-)^I\,\widetilde T_{IM}\,,\\
T_{IM}^c &=& -\widetilde T_{IM}^c\,,\\
\Sigma &=& \widetilde T^\ell_{00}\,,
\eqa
and for $I>0$ and $M\ge 0$
\beqa
T^\ell_{IM}&=&(-)^I\,\frac{1+\delta_{M0}}{2}\,
\Big(\widetilde T^\ell_{IM-}-\widetilde T^\ell_{IM+}\Big)\,,\\
T^\ell_{I-M}&=&-\frac{1+\delta_{M0}}{2}\,
\Big(\widetilde T^\ell_{IM-}+\widetilde T^\ell_{IM+}\Big)\,.
\eqa

\section{Discussion of Results}
\label{discuss}
\begin{figure}
\centerline{%
\epsfxsize=10.0cm
\epsffile{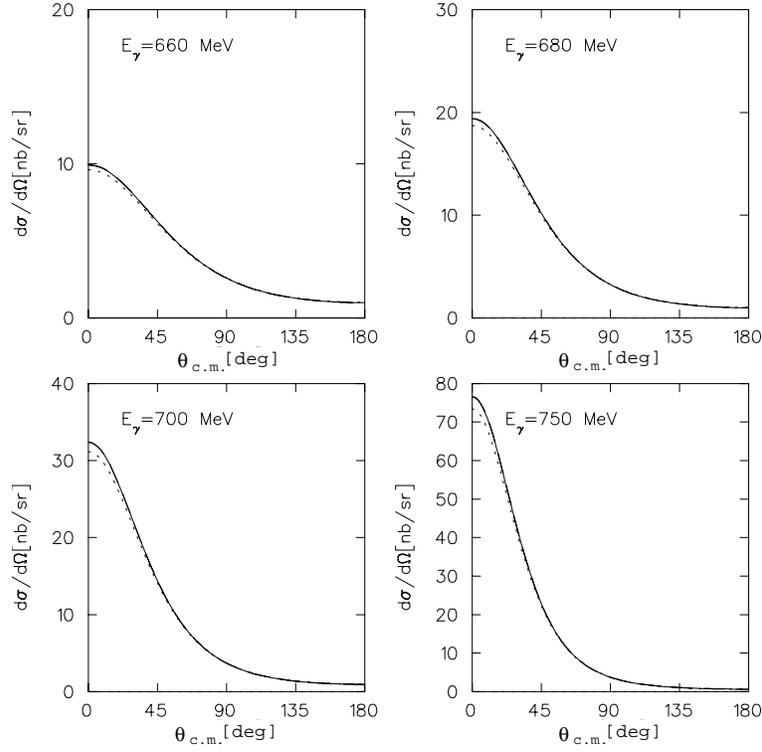}}
\vspace*{.3cm}
\caption{\label{resonances} Differential cross section of $d(\gamma,\eta)d$ 
including the resonance graphs only. Notation of the curves:
 dotted: $S_{11}(1535)$, 
 dashed: $S_{11}(1535)$ + $D_{13}(1520)$,
 full: $S_{11}(1535)$ + $D_{13}(1520)$ + $P_{11}(1440)$.}
\end{figure}

We will begin the discussion of our results by considering the influence of 
the various ingredients on the differential cross section. In 
Fig.~\ref{resonances} we show the resonance contributions at four 
representative photon energies between threshold and the maximum, starting 
with the $S_{11}(1535)$ and consecutively adding the 
$D_{13}(1520)$ and $P_{11}(1440)$ resonances. One readily notices the 
overwhelming dominance of the $S_{11}(1535)$ while the effect of adding the 
$D_{13}(1520)$ is barely seen and the $P_{11}(1440)$ is negligible. 
This result
is in accordance with \cite{GK1} and it is also obvious because of
the small couplings of the $\eta$-meson to the other two resonances. However, 
that their role in combination with rescattering will be correspondingly small 
cannot be inferred at all as long as $\pi$-meson exchange is included 
as is discussed below. 

\begin{figure}
\centerline{%
\epsfxsize=10.0cm
\epsffile{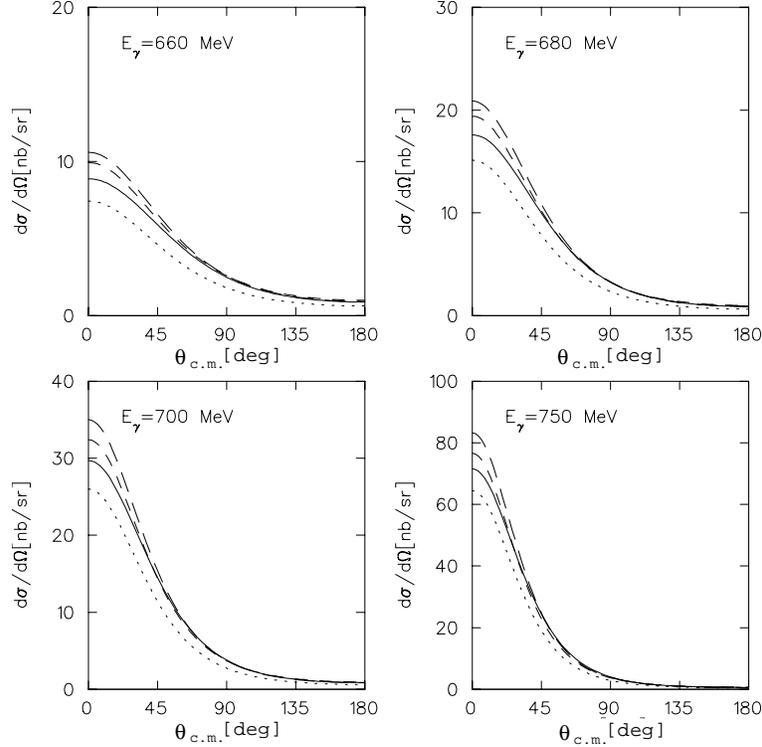}}
\vspace*{.3cm}
\caption{\label{born} Effects of the Born terms on the differential cross 
section of $d(\gamma,\eta)d$. Notation of the curves:
short-dashed: direct resonant graphs 
$S_{11}(1535)$ + $D_{13}(1520)$ + $P_{11}(1440)$, 
long-dashed: + direct and crossed 
nucleonic graphs, including the two unconnected graphs,
dotted: + Z-graphs,
full: + $\omega$-meson contribution = IA\@.}
\end{figure}

The influence of the Born terms are shown in Fig.~\ref{born}. Comparing the
short dashed with the solid curves, one notices that the overall contribution
of the Born terms to the unpolarized differential cross section is rather
small, although the separate contributions like the Z and vector meson graphs
of Fig.~\ref{overviewborn} show very large effects separately, 
but tend to cancel each
other, in agreement with \cite{EB}. Without the vector meson graphs there
would be a sizeable Born contribution.  In summary, only in the very forward
direction one finds a small reduction of a few percent from the Born terms.

\begin{figure}
\centerline{%
\epsfxsize=9.5cm
\epsffile{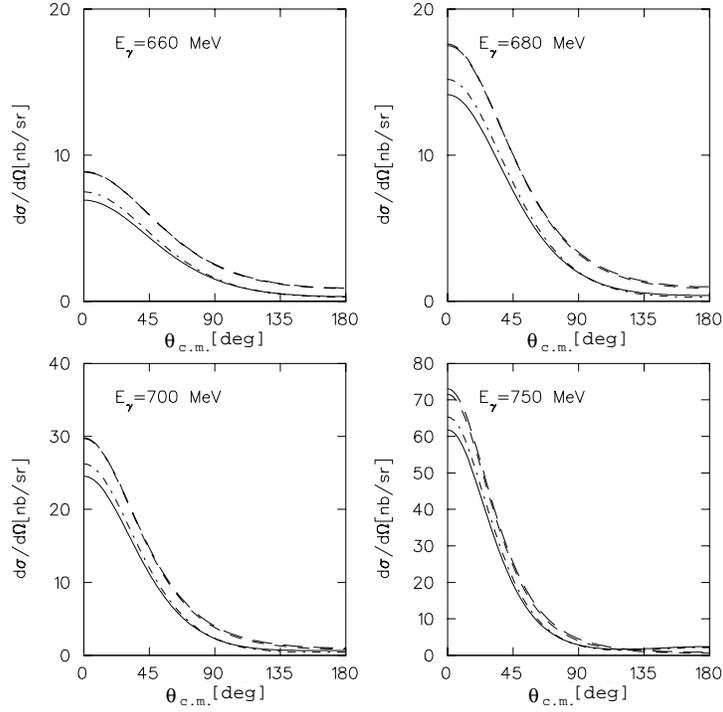}}
\vspace*{.3cm}
\caption{\label{resc1} Effect of the different static rescattering mechanisms
 involving the $S_{11}$ resonance on the differential cross section of 
$d(\gamma,\eta)d$. Notation of the curves:
short-dashed: IA,
long-dashed: IA + static transition 
 $NS_{11}\leftrightarrow{}NN$ rescattering,
dash-dotted: IA + static $NS_{11}\leftrightarrow{}NS_{11}$ rescattering,
full: IA + both $NS_{11}$ rescattering contributions.}
\end{figure}

As next we will discuss the influence of two-body mechanisms, like 
rescattering and MEC, and begin with rescattering taking first the 
purely static approach. The hadronic rescattering is built upon 
a one-boson exchange mechanism as described in Sect.~\ref{deut}, starting 
for the $NN$ channel from a realistic potential, here the Bonn OBEPQ-B
\cite{Mach1}. 
The effect of the various channels are shown in Fig.~\ref{resc1}. 
The particle-interchanging interaction 
$S_{11}N\leftrightarrow{}NS_{11}$ clearly dominates the process,
the pure transition potential $NS_{11}\leftrightarrow{}NN$ shows very small
effects. But in the combination of both the genuine transition potential
shows effects under forward and backward angles. As has been reported 
already in \cite{RA}, the total effect of static rescattering for 
the coupled $NN$-$NS_{11}$ configurations leads  to a sizeable reduction 
of the differential cross section except at the highest energy where one 
notices a slight increase around backward angles.

\begin{figure}
\centerline{%
\epsfxsize=10.0cm
\epsffile{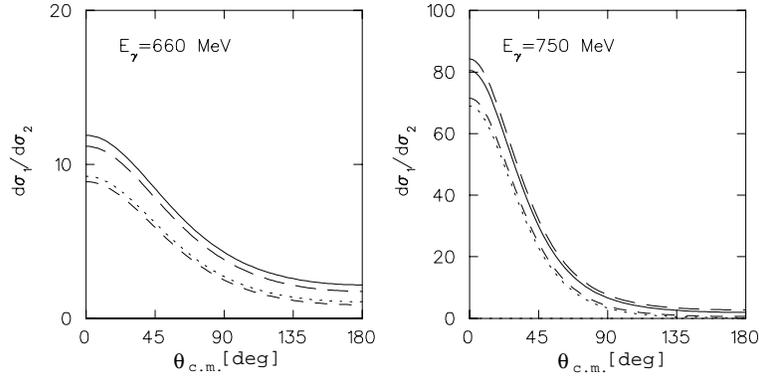}}
\vspace*{.3cm}
\caption{\label{diff_pieta} Effect of the retarded $\pi$- and $\eta$-exchange 
rescattering on the differential cross section of $d(\gamma,\eta)d$. 
Notation of the curves:
short-dashed: IA,
long-dashed: IA + $NN$ rescattering + retarded 
 $\pi$-exchange $NS_{11}$ rescattering,
dotted: IA + $NN$ rescattering + retarded 
 $\eta$-exchange $NS_{11}$ rescattering,
full: IA + $NN$ rescattering + both retarded $NS_{11}$ rescattering
  contributions.}
\end{figure}

However, if one switches on retardation, these findings no longer hold.
Considering first the $NS_{11}$ rescattering contribution, the effect changes
its sign, and one obtains a sizeable increase of the differential cross
section. The reason for this different behavior of static versus retarded 
interaction lies in the fact that
the meson-$NN$ propagator is always negative in the static case (see 
Eq.~(\ref{stat-meson-baryon-prop})), while in the retarded case it is positive 
at low momenta $\vec p$ and $\vec p^{\,\prime}$ according to 
(\ref{meson-baryon-prop}). Thus in this important region of momenta, the 
static and the retarded interactions have opposite sign resulting in 
the noted opposite effect. 
However, the rescattering contributions of the other resonances,
which turn out to be of similar size although somewhat smaller, interfere
destructively with the contributions of the $NS_{11}$ rescattering, so that
one finally ends up with a smaller increase of the differential cross section
which, however, is still noticeable at 90$^\circ$ and for larger angles.  Thus
the lighter resonances become more important via rescattering than their role
in the IA, so that their effect on the differential cross
section is comparable to the $S_{11}(1535)$. But for higher energies their
influence decreases and the rescattering process is dominated by the
$S_{11}(1535)$, as one already notices in the differential cross section for
$E_\gamma=700\;\mbox{MeV}$ at $\theta=90^\circ$.  In view of what has been
said about the dominance of $\eta$-exchange in the rescattering contribution
in \cite{HR} we have evaluated the separate rescattering contributions from
$\pi$- and $\eta$-exchange for two energies, one closer to threshold and the
other near the maximum of the total cross section. The results are presented
in Fig.~\ref{diff_pieta}. One readily notices the dominance of $\pi$-exchange
whereas $\eta$-exchange plays only a minor role although a nonnegligible one.
Furthermore, while near threshold both contributions interfere constructively
they exhibit a destructive interference at higher energies. 

\begin{figure}
\centerline{%
\epsfxsize=10.0cm
\epsffile{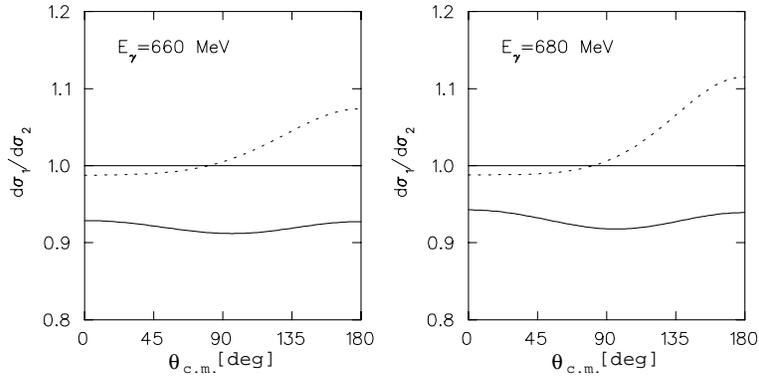}}
\vspace*{.3cm}
\caption{\label{MEC} Relative effect of MEC operators on the differential 
cross section of $d(\gamma,\eta)d$.
Notation of the curves:
dotted: ratio of IA + static $\pi$- and $\eta$-MEC to IA,
full: ratio of IA + all retarded rescattering mechanisms + MEC + $RNN[2]$- and 
$RS_{11}N[2]$-graphs to IA + all retarded rescattering mechanisms.}
\end{figure}

The effect of the pure MEC operators added to the IA is shown in
Fig.~\ref{MEC} as a ratio. The total pure MEC effect turns out to be very
small at forward angles of the differential cross section, of the order of one
percent reduction, whereas at backward angles they lead to an increase of the
order of about 10 percent at 180$^\circ$ depending somewhat on the energy. The
pure MEC is dominated by the pionic graph, whereas the $\eta$ exchange is
largely suppressed. This is in line with the dominance of the pion in the
rescattering contribution.  A different pattern evolves, if one combines the
MEC with the retarded hadronic rescattering graphs as the full curves in
Fig.~\ref{MEC} demonstrate. The combination of MEC contributions with
rescattering leads to a considerably larger effect, namely an almost isotropic
decrease of the differential cross section by about 5 to 8 percent. The reason
for this different feature, obviously, lies in the shorter ranged structure of
the MEC operators compared to the one-body operators. Thus MEC attain some
importance only if rescattering effects are considered modifying the short and
medium range region.

\begin{figure}
\centerline{%
\epsfxsize=10.0cm
\epsffile{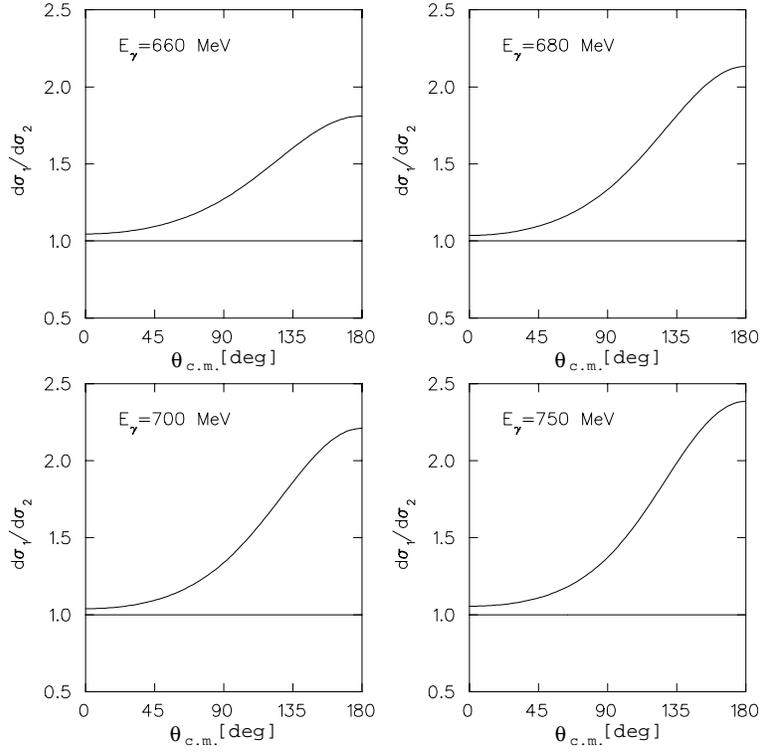}}
\vspace*{.3cm}
\caption{\label{rescTMEC} 
Relative effect of all two-body operators on the differential cross section 
of the coherent reaction.
Notation of the curves:
full: ratio of the complete calculation to IA\@.}
\end{figure}

\begin{figure}
\centerline{%
\epsfxsize=10.0cm
\epsffile{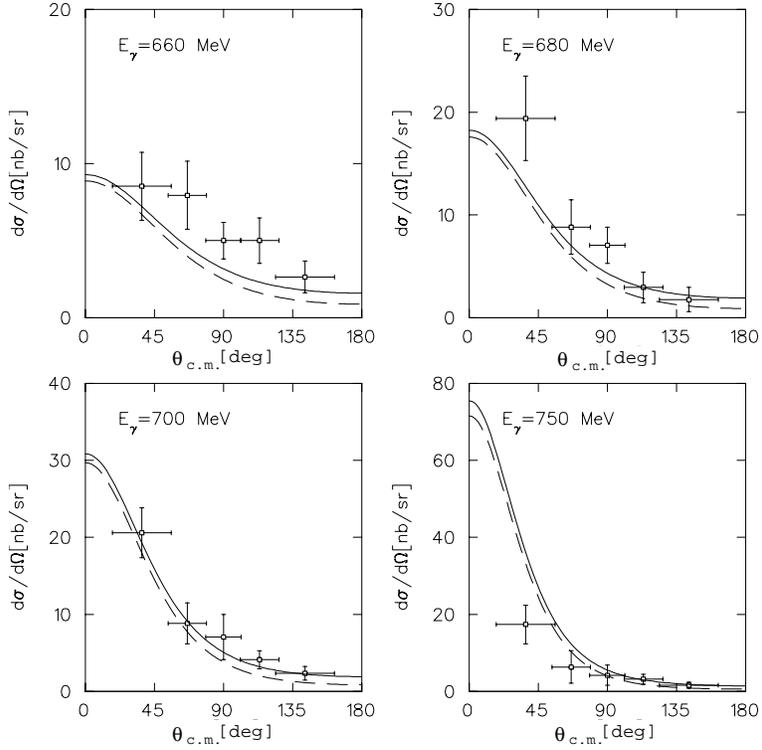}}
\vspace*{.3cm}
\caption{\label{masterdiff}
Summary of all contributions to the differential cross section of 
$d(\gamma,\eta)d$ and comparison to experiment. 
The data points are taken from {\protect\cite{HoffR}}.
Notation of the curves:
dashed: IA,
full: complete calculation, i.e., IA + MEC + retarded rescattering,
including the combination of MEC and rescattering.} 
\end{figure}

The effect of all two-body operators on the differential cross section is
shown in Fig.~\ref{rescTMEC} as a ratio with respect to the pure IA\@. At
forward angles one notes a small increase of a few percent, but the increase
gains steadily with larger angles yielding at 180$^\circ$ an enhancement of
by a factor of about two. But in view of the strong forward peaking of the
differential cross section, the overall effect seems to be quite small.
However, this is misleading because the forward region is suppressed in the
total cross section, so that in fact a sizeable increase remains as is
discussed below. In Fig.~\ref{masterdiff} we compare our results also with the
experimental data of \cite{HoffR}. The description of the data is quite
satisfactory, although the experimental errors are quite large, and for a more
stringent test of the theory data of higher accuracy is needed. We furthermore
would like to emphasize, that we did not use this data to fit any of our model
parameters.  Also we would like to stress the fact, that the
ratio of $|A_s/A_p|\approx 0.2$ extracted previously from the coherent
reaction is compatible with our model. But due to the complex phase relation
we can also reproduce the ratio of the elementary resonant cross
sections $(\sigma_n/\sigma_p)\approx 2/3$ extracted from the incoherent
reaction.

\begin{figure}
\centerline{%
\epsfxsize=9.0cm
\epsffile{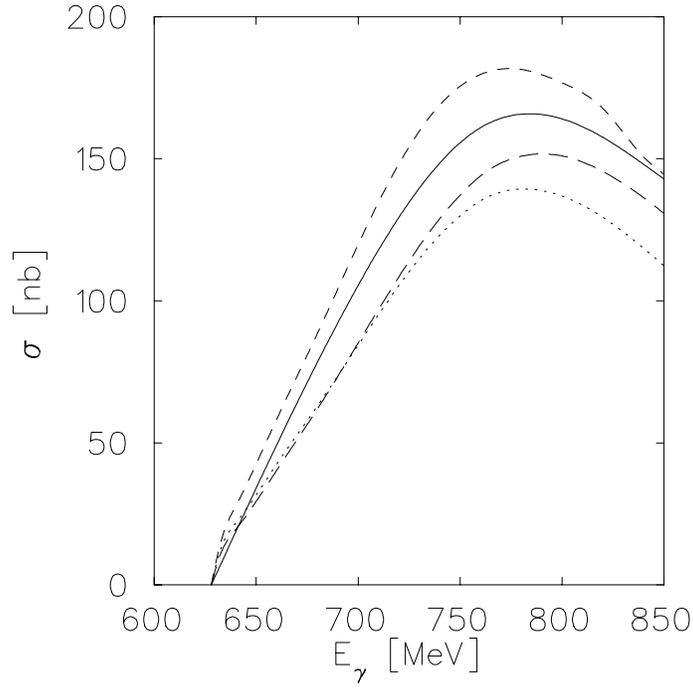}}
\vspace*{.3cm}
\caption{\label{totdiff} Total cross section of the coherent
reaction $\gamma{}d\rightarrow\eta{}d$ for energies up to
$E_\gamma^{lab}=850\;\mbox{MeV}$.
Notation of the curves:
dotted: pure resonance contribution, 
long-dashed: IA,
short-dashed: IA + retarded first order rescattering,
full: complete calculation, i.e., 
 IA + all retarded rescattering contributions + MEC + RNN[2] + RXN[2].}
\end{figure}

The overall effect of two-body mechanisms can be seen more clearly in the
total cross section as shown in Fig.~\ref{totdiff}. They are quite sizeable 
and account for an overall increase which even in the maximum amounts to 
about $10\,\%$ slightly shifting the maximum to lower energies. We 
furthermore show in Fig.~\ref{totdiff} also the result of a rescattering 
treatment in first order replacing the $T$-matrix by the potential $V$. 
Obviously, such an approximate calculation overestimates the rescattering 
effects grossly in agreement with findings in \cite{FiA97}.

\begin{figure}
\centerline{%
\epsfxsize=10.0cm
\epsffile{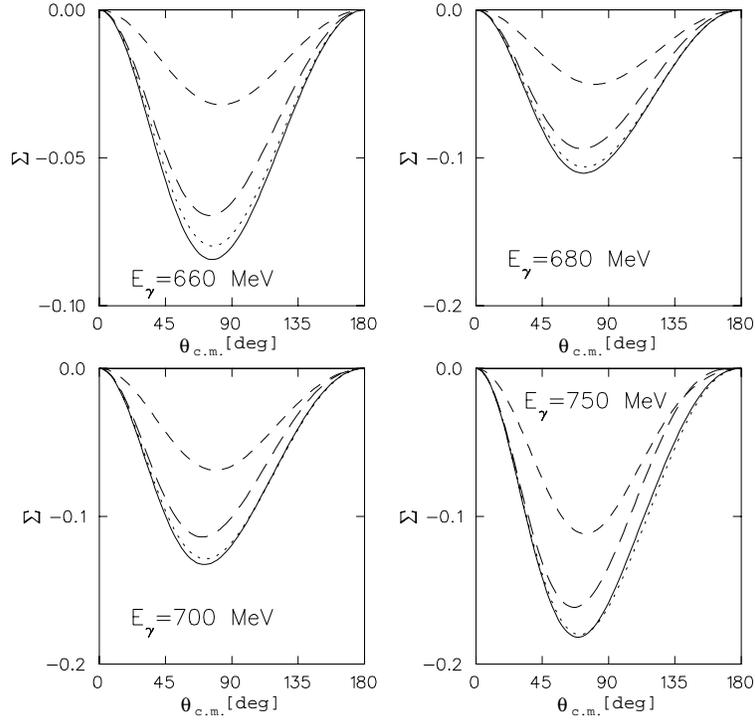}}
\vspace*{.3cm}
\caption{\label{sigma1} Linear photon asymmetry of $d(\gamma,\eta)d$ at 
various photon energies. 
Notation of the curves:
short-dashed: pure resonant contribution,
long-dashed: IA,
dotted: IA + retarded rescattering,
full: complete calculation.}
\end{figure}

\begin{figure}
\centerline{%
\epsfxsize=10.0cm
\epsffile{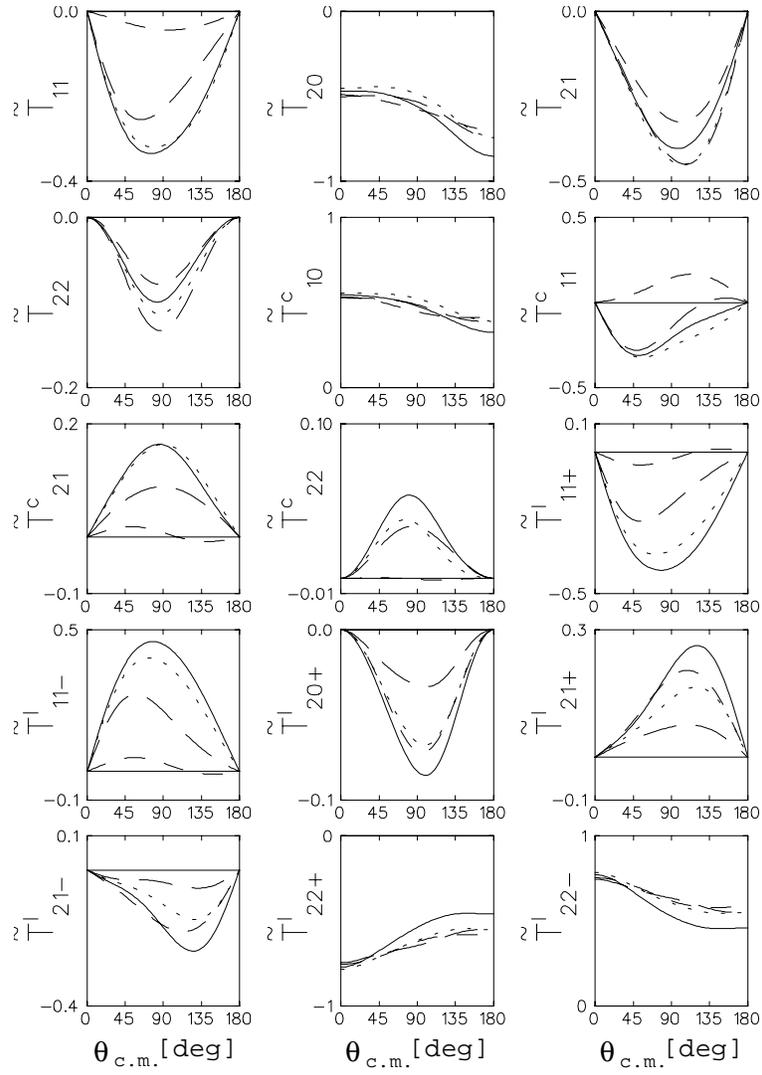}}
\vspace*{.3cm}
\caption{\label{polobs} Target and beam-target polarization observables for
$d(\gamma,\eta)d$ at $E_\gamma^{lab}=700\;\mbox{MeV}$.
Notation of the curves as in Fig.~\protect\ref{sigma1}.}
\end{figure}

Finally, we would like to discuss the polarization observables which usually
are more sensitive to dynamical effects. In Fig.~\ref{sigma1} the various
effects on the linear photon asymmetry $\Sigma$ are presented.  As one
notices, two-body effects are comparably small although not negligible.  It is
interesting that a sizeable amount of the asymmetry stems from the Born terms
being near threshold even larger than the resonance contribution.  The latter,
however becomes more important at the higher energies.  Thus the measurement
of the $\Sigma$-asymmetry would offer the possibility to test whether the
choice of the background terms in the present model is realistic. Target and
beam-target asymmetries are shown in Fig.~\ref{polobs} for one energy, 700
MeV. A quick glance reveals that the different contributions from resonance,
Born, rescattering and MEC manifest themselves in quite different ways in the
various observables. The resonance contribution dominates in $\widetilde
T_{20}$, $\widetilde T_{10}^c$, and $\widetilde T_{22 \pm}^\ell$, the other
contributions being of minor importance. Large Born contributions are found in
$\widetilde T_{11}$, $\widetilde T_{11}^c$, where it leads even to a sign
change, and in $\widetilde T_{11\pm}^\ell$. These observables exhibit also
sizeable to large effects from rescattering, and in addition also in
$\widetilde T_{21\pm}^\ell$. Finally, noticeable effects from MEC can be seen
in $\widetilde T_{22}^c$, $\widetilde T_{11\pm}^\ell$, $\widetilde
T_{20}^\ell$, and $\widetilde T_{21\pm}^\ell$. Thus, a measurement of
polarization observables clearly poses a more detailed test of the underlying
model.

\section{Conclusion}
\label{zusf}

In conclusion we may state that two-body operators give significant
contributions to the total and differential cross section of coherent
$\eta$-meson photoproduction on the deuteron.
Thus these have to be considered in a
detailed comparison with experimental data. With respect to hadronic
rescattering the retarded OBE potentials yield small effects of a few percent
in the forward direction of the differential cross section,
 which, however, largely increase at more backward
angles up to about $100~\%$.
Thus the total cross section shows an increase between $25~\%$ closer to
threshold around 680 MeV and $10~\%$ in the maximum.  Static rescattering
operators seem to be generally much too strong (see also \cite{RA}), showing
an opposite effect and thus
should not be employed.  On the electromagnetic side, the pure meson exchange
currents produce small effects at the forward peak of the differential cross
section, and without combining them with hadronic
rescattering terms they are largely negligible in the present model. But in
combination with hadronic rescattering the MEC operators reduce the overall
two-body effects sizeably. For polarization
observables two-body effects turn out to be important, as there are several
observables which are very sensitive to hadronic rescattering and MEC\@.
Measuring these asymmetries poses a challenging task on the experimental side.
The size of two-body effects may become even larger for the
electroproduction process when entering kinematic regions of higher momentum
transfers.  The description of the available coherent data of \cite{HoffR} is
quite good, although -- as mentioned before -- the big experimental error bars
prevent a conclusive comparison with experiment. We clearly need data of
better quality for the coherent photoproduction of $\eta$-mesons on the
deuteron. Then also the theoretical description of the elementary 
photoproduction reaction as well as the hadronic interaction should be 
improved. 

With respect to the discrepancy for the isoscalar photoproduction amplitude
between coherent and incoherent photoproduction of $\eta$-mesons on the
deuteron, we want to emphasize that the conclusions drawn in \cite{HoffR} are
not stringent. 
The problem in extracting the isoscalar amplitude from the incoherent reaction
lies in the fact, that one has no information on the relative phases between
proton and neutron amplitude. Thus the coherent reaction is more reliable for
obtaining direct access to the isoscalar amplitude.
For the future, a consistent model is needed which describes dynamically meson
nucleon interaction and e.m.\ meson production on the nucleon including at
least two-pion channels. Such a model should include all resonances from the
beginning and treat the intermediate meson propagation retarded.

\begin{acknowledgments}
  We would like to thank Dr.\ M.\ Schwamb for reading of the manuscript and
  many helpful hints, and Dr.\ A.\ Fix for various useful discussions. 
\end{acknowledgments}

\begin{appendix}
\section{Vertices}
\label{vertices}
Here we list the non-relativistic vertices of our model.
The meson emission vertices of the resonances are given by
\beqa
 v^\dag_{B N S_{11}} &=& ig_{BNS_{11}} \tau_B, \\
v^\dag_{B N P_{11}} &=& i \frac{g_{BNP_{11}}}{2m^0_{P_{11}}} 
 \tau_B \vec{\sigma}\cdot\vec{k}, \\
v^\dag_{BND_{13}} &=& 
i \alpha_{D_{13}} \frac{g_{BN{}D_{13}}}{m_\pi} \tau_B \,
\vec{\sigma}_{NN}\!\cdot\!\vec{k}
\vec{\sigma}_{ND_{13}}\!\cdot\!\vec{k}, 
\eqa
where $\tau_B$ denotes the elementary isospin operator
\beq
 \tau_\pi = \tau_\mu^\dag, \quad \tau_\eta = {\mathbb 1}.
\eq
The factor $\alpha_{D_{13}}$ is defined as
\beq
\label{alphad13}
  \alpha_{D_{13}} = 
  \frac{1}{4}\left(\frac{1}{m^0_{D_{13}}}+\frac{1}{m_{N}}\right)
  \approx (16.9\,m_\pi)^{-1}.
\eq

The e.m.\ excitation of the resonances is described by
\beqa
v_{S_{11}\leftarrow\gamma{}N} 
&=& -\frac{\hat{k}_{S_{11}}}{m_N+m^0_{S_{11}}}\omega_\gamma(q)\vec{\sigma}\,,\\
v_{P_{11}\leftarrow\gamma{}N} &=& \frac{1}{2 m^0_{P_{11}}} \hat{k}_{P_{11}}
 i\vec{\sigma}\times\vec{k}\,, \\
v_{D_{13}\leftarrow\gamma{}N} &=&
\frac{\hat{k}_{D_{13}}}{2m_N}
\,\omega_\gamma(q)
\vec{\sigma}_{D_{13}N}\,.
\eqa

Within the framework of time ordered perturbation theory we derived the
following expression for the current operator associated with the 
vector meson graphs, using vector coupling only
\beqa
\vec{\jmath}_{BV\gamma} &=& -\tau_{BV\gamma}\frac{e\lambda_{BV\gamma}}{m_B}
g_V \frac{G^V(W)}{2\omega^V_0(k)} 
\Big\{\frac{\vec{k}_V^2}{m_V^2}\vec{k}_V\times\vec{q}
-\frac{\omega^V_0(k_V)}{m_V^2}\frac{1}{2m_N}(\vec{p}+\vec{p}^{\,\prime})\cdot\vec{k}_V
\,\vec{k}_V\times\vec{q} \nonumber \\ && 
+\frac{\omega_\gamma(q)}{2m_N}\vec{k}_V\times\left((\vec{p}+\vec{p}^{\,\prime})+i\vec{\sigma}\times\vec{k}_V
\right)\Big\},
\eqa
where $\vec{k}_V=\vec{q}-\vec{k}$ is the momentum of the vector meson. 
$G^V(W)$ is the elementary propagator of the intermediate state and 
contains two time orderings
\beqa
 G^V(W) &=& \frac{1}{W-e_N(p)-\omega^V_0(\vec{q}-\vec{k}\,)-\omega_k
 +i\epsilon} 
+\frac{1}{W-e_N(p^\prime)-\omega^V_0(\vec{q}-\vec{k}\,)
 -\omega_\gamma(q) + i\epsilon}.
\eqa
In the deuteron this propagator is slightly more complicated, but develops no
singularities.
The operator $\tau_{BV\gamma}$ is the isospin part of the current. 
For the various physical channels one gets
\beq
\begin{array}{ll}
\tau_{\eta\omega\gamma}={\mathbb 1}, &\quad 
  \tau_{\pi\omega\gamma}=\delta_{\mu,0}, \\
\tau_{\eta\rho\gamma}=\tau_0, &\quad 
  \tau_{\pi\rho\gamma}=\tau_\mu^\dag.
\end{array}
\eq
For the coherent reaction on the deuteron only the $\eta\omega\gamma$ graph
contributes.

\section{Transition Potentials}
\label{transpot}
In this appendix we list  the hadronic transition
potentials. Note that the number of transition potentials increases
quadratically with the number of channels, i.e., in our case we have considered
besides the $NN$-interaction 3 diagonal potentials, 
and 6 genuine transition potentials:
\beqa
 V_{NS_{11}\leftarrow{}NS_{11}} &=&{\Omega}_B
  \frac{g_{BNS_{11}}^2}{(2\pi)^3 2\omega_k}
    \; \left(G_0^{BNN}(W)+G_0^{BS_{11}S_{11}}(W)\right) 
  + (1\leftrightarrow{}2), \\
 V_{ND_{13}\leftarrow{}ND_{13}} &=&{\Omega}_B
  \left(\frac{g_{BND_{13}} \alpha_{D_{13}}}{m_\pi}\right)^2
  \frac{1}{(2\pi)^3 2\omega_k}
  \vec{\sigma}_{RN}(2)\cdot\vec{k}\,
  \vec{\sigma}_{NN}(2)\cdot\vec{k} \nonumber \\ && \times 
  \vec{\sigma}_{NN}(1)\cdot\vec{k}\,
  \vec{\sigma}_{NR}(1)\cdot\vec{k} 
  \; \left(G_0^{BNN}(W)+G_0^{BD_{13}D_{13}}(W)\right) 
 + (1\leftrightarrow{}2), \\
 V_{NP_{11}\leftarrow{}NP_{11}} &=&{\Omega}_B
  \left(\frac{g_{BNP_{11}}}{2m^0_{P_{11}}}\right)^2
  \frac{1}{(2\pi)^3 2\omega_k}
  \vec{\sigma}_1\cdot\vec{k}\,
  \vec{\sigma}_2\cdot\vec{k} 
    \; \left(G_0^{BNN}(W)+G_0^{BP_{11}P_{11}}(W)\right) 
+ (1\leftrightarrow{}2),\\
 V_{NS_{11}\leftarrow{}ND_{13}} &=&{\Omega}_B
  \frac{g_{BNS_{11}}g_{BND_{13}}}{(2\pi)^3 2\omega_k}\frac{\alpha_{D_{13}}}{m_\pi}
    \vec{\sigma}_{NN}(1)\cdot\vec{k}\,
    \vec{\sigma}_{NR}(1)\cdot\vec{k}  
 \left(G_0^{BNN}(W)+G_0^{BS_{11}D_{13}}(W)\right)
 + (1\leftrightarrow{}2), \\
 V_{NS_{11}\leftarrow{}NP_{11}} &=&-{\Omega}_B
  \frac{g_{BNS_{11}}g_{BNP_{11}}}{(2\pi)^3 2\omega_k}
    \frac{\vec{\sigma}_1\cdot\vec{k}}{2m^0_{P_{11}}}
    \; \left(G_0^{BNN}(W)+G_0^{BP_{11}S_{11}}(W)\right) 
 + (1\leftrightarrow{}2), \\
 V_{ND_{13}\leftarrow{}NP_{11}} &=&-{\Omega}_B
  \frac{g_{BND_{13}}g_{BNP_{11}}}{(2\pi)^3 2\omega_k}%
  \frac{\alpha_{D_{13}}}{m_\pi}%
    \vec{\sigma}_{RN}(2)\cdot\vec{k}\,
    \vec{\sigma}_{NN}(2)\cdot\vec{k} 
    \frac{\vec{\sigma}_{NN}(1)\cdot\vec{k}}{2m^0_{P_{11}}} \nonumber \\
   &&\times \left(G_0^{BNN}(W)+G_0^{BP_{11}D_{13}}(W)\right) 
 + (1\leftrightarrow{}2), \\
 V_{NS_{11}\leftarrow{}NN} &=&{\Omega}_B
  \frac{g_{BNS_{11}}g_{BNN}}{(2\pi)^3 2\omega_k}
    \frac{\vec{\sigma}(2)\cdot\vec{k}}{2m_{N}}
    \; \left(G_0^{BNN}(W)+G_0^{BNS_{11}}(W)\right) 
 + (1\leftrightarrow{}2), \\ 
 V_{ND_{13}\leftarrow{}NN} &=&{\Omega}_B
  \frac{g_{BND_{13}}g_{BNN}}{(2\pi)^3 2\omega_k}
  \frac{\alpha_{D_{13}}}{m_\pi}
    \vec{\sigma}_{RN}(2)\cdot\vec{k}\,
    \vec{\sigma}_{NN}(2)\cdot\vec{k} 
    \frac{\vec{\sigma}_{NN}(1)\cdot\vec{k}}{2m_{N}} 
 \left(G_0^{BNN}(W)+G_0^{BND_{13}}(W)\right)
  + (1\leftrightarrow{}2), \\
 V_{NP_{11}\leftarrow{}NN} &=& {\Omega}_B
 \frac{g_{BNP_{11}}g_{BNN}}{4 m^0_{P_{11}} m_N}\frac{1}{(2\pi)^3 2\omega_k}
    \vec{\sigma}(1)\cdot\vec{k}\,\vec{\sigma}(2)\cdot\vec{k}
    \; \left(G_0^{BNN}(W)+G_0^{BNP_{11}}(W)\right) 
  + (1\leftrightarrow{}2).
\eqa
\end{appendix}

\newpage

\begin{table}
\begin{center}
\caption{\label{BTfacs}
Bare resonance masses and the
relation between the hadronic couplings in the Bennhold-Tanabe ansatz
 $g^{[BT]}_X$ and the $g_X$ of the Lagrangians in 
Eqs.~(\protect\ref{lbns11})--(\protect\ref{lbnd13}). 
The factor $\alpha_{D_{13}}$ is defined in Eq.~(\protect\ref{alphad13}).}
\begin{tabular}{cccc}
  & $P_{11}(1440)$ & $D_{13}(1520)$ &  $S_{11}(1535)$\\
\hline
 $m_R^0$ [MeV] & $1672.0$ & $1543.7$ & $1555.6$ \\
\hline
 $g_{\pi{}NR}=g^{[BT]}_{\pi{}NR}\times$ 
    & $\frac{2m_N}{m_\pi\sqrt{12\pi}}$ & 
$\alpha_{D_{13}}^{-1}\frac{1}{m_\pi\sqrt{4\pi}}$ & $\frac{1}{\sqrt{12\pi}}$\\
 $g_{\eta{}NR}=g^{[BT]}_{\eta{}NR}\times$ 
   &  $\frac{2m_N}{m_\eta\sqrt{4\pi}}$ 
   & $\sqrt{\frac{3}{4\pi}}\alpha^{-1}_{D_{13}}\frac{m_\pi}{m_\eta^2}$
   & $\frac{1}{\sqrt{4\pi}}$ 
\end{tabular}
\end{center}
\end{table}

\begin{table}
\begin{center}
\caption{\label{gefitte}
Parameters of the effective two-pion widths.}
\renewcommand{\arraystretch}{1.40}
\begin{tabular}{cccc}
  & $P_{11}(1440)$ & $D_{13}(1520)$ & $S_{11}(1535)$ \\
\hline
$\gamma^{\pi\pi}_R$ [MeV] & 80.3 &  24.2 & 4.3 \\
\end{tabular}
\renewcommand{\arraystretch}{1.00}
\end{center}
\end{table}

\begin{table}
\begin{center}
\caption{%
\label{tablecoupl}
Parameters of the  effective e.m.\ resonance couplings in 
Eqs.~(\protect\ref{para1}) and (\protect\ref{para2}).}
\begin{tabular}{crrrcrrr}
 & $P_{11}$ & $D_{13}$ & $S_{11}$& 
 & $P_{11}$ & $D_{13}$ & $S_{11}$\\
\hline
$a^{(0)}$  &  3.1342       & --1.9376 & 0.3176    &
$a^{(1)}$ & 4.2067 & 4.1469&7.0066 \\
$b^{(0)}$  & --0.1103        &2.2500 & --0.0029   &
$b^{(1)}$ & --0.2762 &--0.4408& 0.0090\\
$c^{(0)}$  & --0.2119       &--0.3790& --0.0055   & 
$c^{(1)}$ & --0.1140 &--0.2615&--0.6776 \\
$d^{(0)}$  & --0.0406        &--0.1368& --0.0007    & 
$d^{(1)}$ &--0.0328  &--0.0383& --0.0016\\
$e^{(0)} $  &  0.0165        & 0.0290 &  0.0007   &
$e^{(1)}$ & 0.0056 &0.0257&  0.0145\\
$f^{(0)}$ [rad]  &  --4.0157& 0.8250 & --0.0830&
$f^{(1)}$ [rad] & --1.3631 & 0.0139& 0.2389\\
$g^{(0)}$ [rad]  &   2.6610  & --0.8955 &  --0.0270 &
$g^{(1)}$ [rad]& --0.0357 &1.0541  &0.0091\\
$h^{(0)}$ [rad]  & --0.4728  &0.1962 & --0.0228  &
$h^{(1)}$ [rad] &--0.1931  &0.0047&--0.0173 \\
\end{tabular}%
\end{center}
\end{table}

\begin{figure}
\centerline{%
\epsfxsize=9.5cm
\epsffile{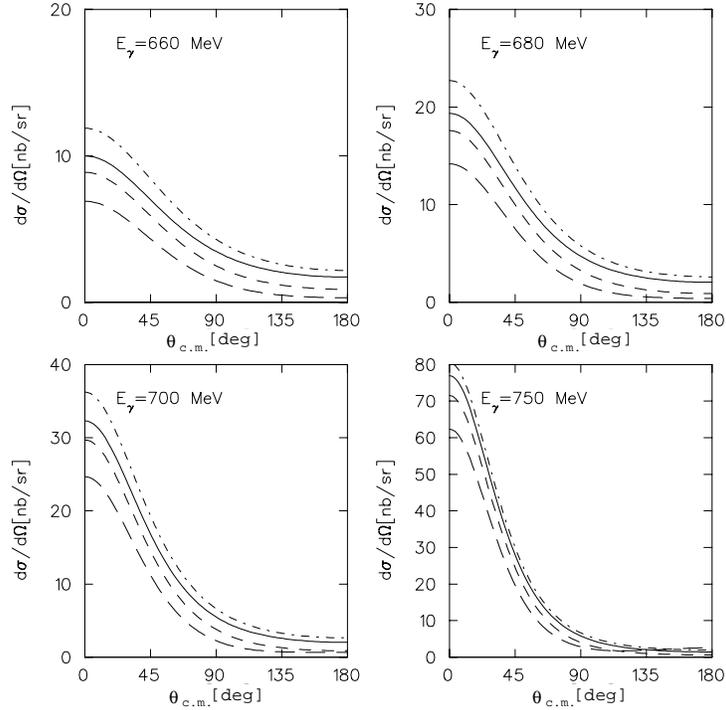}}
\vspace*{.3cm}
\caption{\label{resc2} Effect of the different retarded rescattering 
mechanisms on the differential cross section of $d(\gamma,\eta)d$. 
Notation of the curves:
short-dashed: IA,
long-dashed: IA + $NN$ rescattering + static $NS_{11}$-rescattering,
dash-dotted: IA + $NN$ rescattering + retarded $NS_{11}$-rescattering,
full: IA + all retarded rescattering contributions, i.e., including the higher
resonances but without MEC\@.} 
\end{figure}

\end{document}